\documentclass[traditabstract]{aa}

\usepackage{graphicx}
\pdfoutput=1
\usepackage{amsmath}
\usepackage{txfonts}
\usepackage[round]{natbib}
\usepackage[usenames,dvipsnames]{color}
\usepackage{enumerate}
\newcommand{\binc}{\texttt{binary\_c}}
\newcommand{\Msun}{M_{\odot}}
\newcommand{\Mprim}{M_{1,\mathrm{i}}}
\newcommand{\Menvmin}{M_{\mathrm{env,min}}}
\newcommand{\Msec}{M_{2,\mathrm{i}}}
\newcommand{\Mpmz}{M_{\mathrm{PMZ}}}

\newcommand{\Rsun}{\mathrm{R}_{\odot}}
\newcommand{\Hy}{\mathrm{H}}
\newcommand{\C}{\mathrm{C}}
\newcommand{\Cth}{^{13}\mathrm{C}}

\newcommand{\Ox}{\mathrm{O}}

\newcommand{\Ne}{\mathrm{Ne}}
\newcommand{\Na}{\mathrm{Na}}

\newcommand{\Fe}{\mathrm{Fe}}
\newcommand{\Ge}{\mathrm{Ge}}
\newcommand{\Sr}{\mathrm{Sr}}

\newcommand{\Ba}{\mathrm{Ba}}
\newcommand{\Pb}{\mathrm{Pb}}
\newcommand{\Eu}{\mathrm{Eu}}
\newcommand{\X}{\mathrm{X}}
\newcommand{\Y}{\mathrm{Y}}
\newcommand{\El}{\mathrm{El}}
\newcommand{\ls}{\mathrm{ls}}
\newcommand{\hs}{\mathrm{hs}}

\newcommand{\logg}{\log g}
\newcommand{\loggunits}{\log_{10}(g/\mathrm{cm}\,\mathrm{s}^{-2})}
\newcommand{\Teff}{T_{\mathrm{eff}}}
\newcommand{\Pin}{P_{\mathrm{i}}}
\newcommand{\Pf}{P_{\mathrm{f}}}
\newcommand{\Porb}{P_{\mathrm{orb}}}
\newcommand{\chisq}{\chi^2}
\newcommand{\chimin}{\chi^2_{\mathrm{min}}}
\newcommand{\redchi}{\chi^2_{\nu}}
\newcommand{\aCE}{\alpha_{\mathrm{CE}}}
\newcommand{\Macc}{M_{\mathrm{acc}}}
\newcommand{\Ms}{M_*}
\newcommand{\Mc}{M_{\mathrm{c}}}
\begin{document}
   \title{Carbon-enhanced metal-poor stars: a window\\ on AGB nucleosynthesis and binary evolution (I)}
   \subtitle{Detailed analysis of $15$ binary stars with known orbital periods.}
   \titlerunning{CEMP stars: a window on AGB nucleosynthesis and binary evolution (I)}

   \author{C. Abate
          \inst{1,3}
          \and
          O. R. Pols \inst{1} 
		  \and
		  A. I. Karakas \inst{2}
          \and
          R. G. Izzard \inst{3}
          }

   \institute{Department of Astrophysics/IMAPP, Radboud University Nijmegen, P.O. Box 9010, 6500 GL Nijmegen, The Netherlands
         \and
         Research School of Astronomy \& Astrophysics, Mount Stromlo Observatory, Weston Creek ACT 2611, Australia
%         \email{akarakas@mso.anu.edu.au}
         \and
             Argelander Institut f\"ur Astronomie, Auf dem H\"ugel 71, D-53121 Bonn, Germany\\
              \email{cabate@uni-bonn.de}
%              \email{izzard@astro.uni-bonn.de}
             }

   \date{Received ...; accepted ...}

%%%%%%%%%%%%%%%%%%%%%%%%%%%%%  ABSTRACT  %%%%%%%%%%%%%%%%%%%%%%%%%%
 
  \abstract
   {
   %Context
   %{
   AGB stars are responsible for producing a variety of elements, including carbon, nitrogen, and the heavy elements produced in the slow neutron-capture process ($s$-elements). 
   There are many uncertainties involved in modelling the evolution and nucleosynthesis of AGB stars, and this is especially the case at low metallicity, where most of the stars with high enough masses to enter the AGB have evolved to become white dwarfs and can no longer be observed. The stellar population in the Galactic halo is of low mass ($\lesssim 0.85\Msun$) and only a few observed stars have evolved beyond the first giant branch. However, we have evidence that low-metallicity AGB stars in binary systems have interacted with their low-mass secondary companions in the past.
   %}
   %Aims
   %{
   The aim of this work is to investigate AGB nucleosynthesis at low metallicity by studying the surface abundances of chemically peculiar very metal-poor stars of the halo observed in binary systems.
   %}
   %Methods
   %{
   To this end we select a sample of $15$ carbon- and $s$-element-enhanced metal-poor (CEMP-$s$) halo stars that are found in binary systems with measured orbital periods. 
   With our model of binary evolution and AGB nucleosynthesis, we determine the binary configuration that best reproduces, at the same time, the observed orbital period and surface abundances of each star of the sample. The observed periods provide tight constraints on our model of wind mass transfer in binary stars, while the comparison with the observed abundances tests our model of AGB nucleosynthesis.
   %}
   %Results
   %{
   For most of the stars in our sample, we find that an episode of efficient wind mass transfer, combined with strong angular momentum loss, has occurred in the past. In some cases we find discrepancies between the observed and modelled abundances even if we adopt a fine-tuned set of parameters in our binary evolution model. These discrepancies are probably caused by missing physical ingredients in our models of AGB nucleosynthesis and they provide indications of how to improve our knowledge of the process of nucleosynthesis in AGB stars.
   %}
   %Conclusions
   {}
   }

\nopagebreak
\maketitle

%%%%%%%%%%%%%%%%%%%%%%%%%%%  INTRODUCTION  %%%%%%%%%%%%%%%%%%%%%%%%%%%
\section{Introduction}
\label{intro}

In the final nuclear-burning stage of their lives, stars with initial masses between about $0.8\,\Msun$ and $8\,\Msun$ ascend the asymptotic giant branch (AGB). During this phase of evolution the stellar radius and luminosity increase by two to three orders of magnitude and the convective envelope of the star is expelled and enriches the interstellar medium with the products of stellar nucleosynthesis. AGB stars play an important role in our understanding of the origin of the elements \cite[][]{Travaglio2004, Romano2010, Kobayashi2011}.
Nuclear reactions in the interior of AGB stars are responsible for producing a variety of isotopes of elements, such as carbon, nitrogen, fluorine, sodium, magnesium, and also elements heavier than iron produced by the slow neutron-capture process \cite[][]{Busso1999, Herwig2005}.

Despite the importance of AGB stars for the chemical evolution of galaxies, several aspects of their evolution are not well understood. The physics of mixing is poorly constrained and the connected roles of convection, overshooting, rotation, and magnetic fields need to be analysed in detail.
Mass loss in AGB stars is highly uncertain, and most prescriptions in the literature are based on semi-empirical fits to observations of AGB stars in our Galaxy or in the Magellanic Clouds \cite[e.g.][]{VW93, vanLoon2005}.
Other uncertainties include nuclear reaction rates and low-temperature opacities that follow the chemical composition of the AGB model in detail; both are crucial for determining the predicted level of chemical enrichment from an AGB model \cite[e.g.][]{Izzard2007, Marigo2009}.

AGB nucleosynthesis depends on all these physical quantities, and therefore the observation of chemical abundances in AGB stars and their progeny (e.g. post-AGB stars and planetary nebulae) are an important source of information on the physical processes that drive the evolution of these objects \citep{Herwig2005}. However, because the duration of the AGB phase is typically less than $1\%$ of the total stellar lifetime \citep{VW93}, AGB stars are relatively rare compared to stars in earlier evolutionary stages. This is a limit especially at the low metallicity of the Galactic halo, where the stellar population is about ten billion years old and where AGB stars more massive than approximately $0.85\,\Msun$ have already become white dwarfs.

One way to overcome this limit is to look for signatures of AGB nucleosynthesis in halo binary stars. Consider a binary system where, in the long distant past, the primary star ascended the AGB. When this happened, some fraction of its stellar wind was transferred onto the companion, a lower-mass secondary star. If the total mass of the secondary star after the accretion did not exceed approximately $0.85\,\Msun$, this star has not yet become a white dwarf and can in principle still be observed today.

This binary scenario has been invoked to explain the peculiar chemical abundances determined in the carbon-enhanced metal-poor (CEMP) stars widely observed among the very metal-poor (here defined as%
\footnote{[X/Y]$= \log_{10} (N_{\X}/N_{\Y}) - \log_{10} (N_{\X\odot}/N_{\Y\odot})$, where $N_{\mathrm{X,Y}}$ indicates the number density of the elements X and Y, and $\odot$ denotes the abundances in the Sun.} %
$[\mathrm{Fe}/\mathrm{H}]\lesssim-2.0$) stars of the Galactic halo. CEMP stars constitute a significant fraction of metal-poor stars in the halo, between 9\% and 25\% \cite[e.g.][]{Marsteller2005,Frebel2006, Lucatello2006, Carollo2012, Lee2013}, with the CEMP frequency rising with a decrease in metallicity \cite[][]{Carollo2012, Yong2013II, Lee2013}. Most CEMP stars are also enriched in heavy elements produced by slow and rapid neutron-capture processes ($s$-process and $r$-process, respectively). Traditionally the excess of barium in stars is related to $s$-process and the excess of europium to $r$-process \cite[][]{Sneden2008}. \cite{Jonsell2006} indicate as CEMP those very metal-poor stars with observed $[\C/\Fe]>1$ and as CEMP-$s$ the CEMP stars that satisfy the relations $[\Ba/\Fe] > 1$ and $[\Ba/\Eu] > 0$ simultaneously. CEMP-$s$ stars with $[\Eu/\Fe] > 1$ are defined CEMP-$s/r$ stars. Other authors adopt slightly different definitions \cite[e.g.,][]{BeersChristlieb2005, Aoki2007, Masseron2010}.
The binary formation scenario is supported by detection of radial velocity variations in a large number of CEMP-$s$ stars, statistically consistent with the hypothesis that all CEMP-$s$ stars are in binary systems \cite[][]{Lucatello2005b, Starkenburg2014}.

Several authors have used AGB nucleosynthesis models with the aim to reproduce the abundances observed in CEMP-$s$ stars \citep{Stancliffe2008, Stancliffe2009, Bisterzo2009, Masseron2010, Bisterzo2011, Bisterzo2012, Lugaro2012, Placco2013}. In these studies the observed abundances of the CEMP-$s$ stars are directly compared to the outcome of detailed models of AGB nucleosynthesis of various masses and metallicities, with some assumptions to estimate the dilution of the accreted material in the envelope of the secondary star.
In particular, \cite{Bisterzo2012} provide an individual analysis of $94$ CEMP-$s$ stars in the metallicity range $-3.6\le[\Fe/\Hy]\lesssim-1.0$.  \cite{Placco2013} have recently performed a similar analysis of two newly discovered CEMP-$s$ stars for which they had previously determined the surface abundances of $34$ elements. 
\cite{Bisterzo2012} and \cite{Placco2013} conclude that the observed abundances of most CEMP-$s$ stars are consistent with the hypothesis that these stars have accreted mass from an AGB companion, while the models of AGB nucleosynthesis do not agree well with the abundances of CEMP-$s/r$ stars. The physical process that leads to simultaneous enhancement of $s$- and $r$- elements in these stars is still debated.

In this paper we focus our analysis on a sample of $15$ observed CEMP-$s$ binary stars with known orbital periods. We tackle the problem of their formation history by studying the whole dynamical and chemical evolution of the binary systems. The purpose of this work is to study under which conditions our model reproduces at the same time the evolutionary stage, chemical abundances and orbital period of each observed star, and to understand which constraints can be placed on our model of binary evolution and of the nucleosynthesis processes in AGB stars. The measurement of the orbital period, besides implying that these systems have very likely interacted in the past, provides a constraint on the initial orbital separation and hence also on the initial masses of the stars in the binary system. A larger sample of CEMP-$s$ stars without measured period is analysed in our forthcoming paper \cite[][{\it submitted}, hereinafter Paper II]{Abate2015-1}.

The paper is structured as follows. In Sect. \ref{data} we describe our sample of observed CEMP-$s$ stars. In Sect. \ref{det-model} we briefly discuss the evolution of a star in the AGB phase according to detailed models. In Sect. \ref{binc} we summarise the main characteristics of our binary evolution model, we explain how we included therein the results of detailed models of AGB nucleosynthesis and we describe our method to find the best fit of the observed abundances of each star in our sample. Sect. \ref{results} is dedicated to the comparison between the data and the outcome of our model. In Sect. \ref{discussion} we discuss the results while Sect. \ref{conclusions} concludes.

%%%%%%%%%%%%%%%%%%%%%%%%%%%  SECTION 2  %%%%%%%%%%%%%%%%%%%%%%%%%%%
\section{Data sample}
\label{data}
To perform our analysis we select a sample of $11$ binary stars according to the following criteria: 
($i$) measured orbital period; ($ii$) iron abundance $-2.8<[\Fe/\Hy]\le -1.8$; ($iii$) enhanced carbon and barium abundance, respectively $[\C/\Fe]\ge1$ and $[\Ba/\Fe]\ge0.5$.
The restriction on the abundance of iron is motivated because our model of AGB nucleosynthesis is tailored to reproduce the abundances at metallicity $Z=10^{-4}$, roughly corresponding to $[\Fe/\Hy]\approx-2.3$ (see details in Sects. \ref{det-model} and \ref{binc}).
We ignore systems in which only upper or lower limits are available. We add to this sample four more systems: 
CS$22956-028$, in which barium is only weakly enhanced, $[\Ba/\Fe]= 0.38$, but the strontium abundance is $[\Sr/\Fe]= 1.39$; 
CS$29497-034$, which is enhanced in carbon and barium ($[\C/\Fe]= 2.69$, $[\Ba/\Fe]= 2.12$) and has many observed elements but the iron abundance is low, $[\Fe/\Hy]= -2.96$; HD$198269$ and HD$201626$, in which barium has not been measured but the abundances relative to iron of other $s$-elements, e.g. lanthanum, cerium and lead, are enhanced by more than 1 dex. In Table \ref{tab:sample} we list for all $15$ stars in our sample the observed orbital period, $\Porb$, eccentricity, $e$, mass function, $f(\Mc)$ surface gravity, effective temperature, $\Teff$, iron abundance and number of elements observed. 
For every star we collect from the literature the absolute abundance of each element,
\begin{equation}
A_{\X}=12+\log_{10} \frac{N_{\X}}{N_{\Hy}}~~,\label{eq:abund}
\end{equation}
where $N_{\X}$ and $N_{\Hy}$ are the number densities of element X and hydrogen, respectively.
To compute the abundances relative to iron, [X/Fe], we use the solar abundances as determined by \cite{Asplund2009}.

%%%%%%%%%%%%%%%%%%%%% TABLE1 %%%%%%%%%%%%%%%%%%%%%%%%
\begin{table*}
\footnotesize
\caption{Summary of the orbital parameters, surface gravities, temperatures and chemical properties observed in our sample CEMP-$s$ stars.}
\label{tab:sample}
\centering
\begin{tabular}{l c c c c c c c c c c}
\hline
\hline
\hspace{1cm}ID & $\Porb$/d & $e$ & $f(\Mc)/\Msun$ & $\loggunits$ & $\Teff$/K & elem. & $[\Fe/\Hy]$ &  $[\C/\Fe]$ & $[\Ba/\Fe]$ & Reference\\
\hline
BD$+04^{\circ}2466$ & $4593$& $0.286$ & $0.076$  & $1.8\pm0.2$& $5032$	   & 20 & $-2.1$ & $1.3$ & $ 1.6 $ & 19, 26, 27\\
CS22942--019 & $2800$		   & $0.1$     & $0.036$ & $2.2\pm0.4$ & $4967$    & 18 & $-2.7$ & $2.2$ & $1.8$ & 6, 8, 10\\
CS22948--027 & $426.5$  	   & $0.02$    & $0.004$ & $1.8\pm0.4$ & $4800$ 	   & 21 & $-2.5$ & $2.2$ & $2.0$ & 17, 22\\
CS22956--028 & $1290$		   & $0.22$    & $0.076$ & $3.9\pm0.1$ & $6900$ 	   & 11 & $-2.1$ & $1.9$ & $0.4$ & 5, 13\\
CS22964--161A& $252.48$ 	   & $0.66$    & $0.147$ & $3.7\pm0.2$ & $6050$ 	   & 21 & $-2.4$ & $1.6$ & $1.4$ & 25\\
CS22964--161B& $252.48$ 	   & $0.66$    & $0.216$ & $4.1\pm0.4$ & $5850$ 	   & 22 & $-2.4$ & $1.4$ & $1.3$ & 25\\
CS29497--030 & $342$		   & $0$	   & $0.002$ & $4.0\pm0.5$ & $6966$    & 33 & $-2.5$ & $2.4$ & $2.3$ & 5, 13, 16, 18, 23\\
CS29497--034 & $4130$		   & $0.02$    & $0.060$ & $1.65\pm0.15$  & $4850$    & 20 & $-3.0$ & $2.7$ & $2.1$ & 3, 17, 21\\
CS29509--027 & $196$		   & $0.15$    & $0.001$ & $4.2\pm0.1$ & $7050$ 	  &  5 & $-2.1$ & $1.5$ & $1.3$ & 4, 13\\
HD198269   & $1295$ 		 & $0.09$	 & $0.107$ & $1.30\pm0.25$& $4800$  	&  9 & $-2.2$ & $1.7$ & $-   $ & 1, 14, 20\\
HD201626   & $407$  		 & $0$  	 & $0.075$ & $2.25\pm0.25$& $5200$  	& 10 & $-2.1$ & $2.1$ & $-   $ & 1, 14, 20\\
HD224959   & $1273$ 		 & $0.179$   & $0.096$ & $1.95\pm0.25$& $5050$  	& 15 & $-2.1$ & $1.8$ & $2.2$ & 1, 14, 20, 28\\
HE0024--2523 & $3.14$		  & $0$ 	  & $0.049$ & $4.3\pm0.1$ & $6625$  	 & 17 & $-2.7$ & $2.1$ & $1.6$ & 10, 11, 12, 15\\
HE0507--1430 & $446$ 		  & $-$ 	  & $-$ 	& $0.8\pm0.1$  & $4600$ 	  & 5 & $-2.4$ & $2.8$ & $1.3$ & 24, 29\\
LP625--44	 & $\gtrsim4383$   & $-$	   & $-$	 & $2.65\pm0.3$& $5500$ 	  & 31 & $-2.8$ & $2.3$ & $2.8$ & 2, 3, 7, 8, 21\\
\hline
\end{tabular}
\tablebib{
(1)~\citet{Vanture1992b,Vanture1992c};
(2)~\citet{Norris1997};
(3) \citet{Aoki2000};
(4)~\citet{Hill2000};
(5)~\citet{Preston2000};
(6) \citet{Preston2001};
(7) \citet{Aoki2001};
(8) \citet{Aoki2002-1};
(9) \citet{Aoki2002-5};
(10) \citet{Cohen2002};
(11) \citet{Carretta2002};
(12) \citet{Lucatello2003};
(13) \citet{Sneden2003-2};
(14) \citet{VanEck2003};
(15) \citet{Cohen2004};
(16) \citet{Sivarani2004};
(17) \citet{Barbuy2005};
(18) \citet{Ivans2005};
(19) \citet{Jorissen2005};
(20) \citet{Lucatello2005b};
(21) \citet{Aoki2006};
(22) \citet{Aoki2007};
(23) \citet{Johnson2007};
(24) \citet{Beers2007};
(25) \citet{Thompson2008};
(26) \citet{Pereira2009};
(27) \citet{Ishigaki2010};
(28) \citet{Masseron2010};
(29) \citet{Hansen2012};
}
\end{table*}
%%%%%%%%%%%%%%%%%%%%%%%%%%%%%%%%%%%%%%%%%%%%%%%%%%%%%%%%%%

In most of the stars of our sample the abundances determined by different authors are consistent within the observational uncertainties. In these cases we adopt the arithmetic mean of the absolute abundances and the maximum published uncertainty. In star CS$22948-027$ there are large discrepancies between the abundances published by different authors, up to $1$ dex or more. These discrepancies are mostly due to the different atmospheric parameters adopted: \cite{Preston2001} and \cite{Aoki2002-4} find a low surface gravity, $\loggunits=0.8$ and $1.0$ respectively, whereas \cite{Hill2000}, \cite{Barbuy2005} and \cite{Aoki2007} find higher values, $\loggunits=1.8,\,1.8$ and $1.9$, respectively. The most recent abundances published by \cite{Barbuy2005} and \cite{Aoki2007} are obtained from spectra at higher resolution and signal-to-noise ratio compared to previous studies and include corrections for effects due to non-local thermodynamic equilibrium, thus in our study we adopt the average of their abundances. In stars HD$198269$ and HD$201626$ the iron abundances published by \cite{Vanture1992b} are $0.8$ dex higher than the values indicated by \cite{VanEck2003}, although the abundances of most $s$-elements are consistent within the observational uncertainties. In the analysis of these two stars we use the abundances of iron and $s$-elements published by \cite{VanEck2003} that are obtained from spectra at higher resolution. Because the abundances of carbon, nitrogen and oxygen are not provided by \cite{VanEck2003} we adopt the values indicated by \cite{Vanture1992a, Vanture1992b}.

The minimum uncertainty that we assume is $0.1$ dex in the chemical abundances and $\logg$. For the effective temperature we adopt an uncertainty of $100\,$K unless differently stated. Orbital periods and eccentricities of the $15$ binary stars in our sample are mostly published without errors; in Sect. \ref{method} we explain how we deal with this missing information and which parameters we actually use in our study. For stars HE$0507-1430$ and LP$625-44$ no information about the eccentricity or mass function is available. We adopt $\Porb=12\,$years as period of LP$625-44$, although this is in fact a lower limit \citep{Aoki2002-1} and the real period is probably much longer \cite[][]{Hansen2012}.

%%%%%%%%%%%%%%%%%%%%%%%%%%%  SECTION 3  %%%%%%%%%%%%%%%%%%%%%%%%%%%
\section{Nucleosynthesis during the AGB phase}
\label{det-model}
For a detailed review on AGB evolution we refer to \cite{Herwig2005}. Here we summarise the thermal-pulse cycles that take place in AGB stars. A thermal pulse (TP) occurs when the helium shell ignites and expands the outer layers extinguishing the hydrogen shell. Third dredge-up (TDU) may occur after the TP, when the the inner edge of the convective envelope moves inward (in mass) and mixes to the surface the products of internal nucleosynthesis. During the interpulse period hydrogen is burned quiescently and the $s$-process occurs in the intershell region where helium is abundant and ($\alpha$, $n$) reactions can be efficiently activated to produce free neutrons.

In low-mass stars up to $3\Msun$ the main neutron source is the $^{13}\C(\alpha,\,n)^{16}\Ox$ reaction. $\Cth$ is formed in the intershell region by partial mixing of protons from the envelope at the end of a TDU event. The protons are captured by the $^{12}\C$ to form a layer rich in $\Cth$, the ``$\Cth$ pocket''. During the subsequent interpulse period the $\Cth(\alpha,\,n)^{16}\Ox$ reaction takes place in the $\Cth$ pocket and releases free neutrons \citep{Straniero1995}. One of the largest uncertainties in the study of the $s$-process nucleosynthesis is related to the numerical treatment of the $\Cth$ pocket: its mass is essentially a free parameter in AGB models but some constraints can be put by comparison to the chemical composition observed in planetary nebulae, post-AGB stars and CEMP-$s$ stars. We refer to \cite{Busso1999} for a thorough discussion of the nucleosynthesis in AGB stars.

In the models of \citet[K10 hereinafter]{Karakas2010}, computed with the Monash/Mount Stromlo (hereinafter Stromlo) code, the evolution of the stellar structure is computed as a first step and subsequently detailed nucleosynthesis calculations are performed using a post-processing algorithm \cite[][]{Lugaro2004, Karakas2007}. In the post-processing algorithm, protons are mixed into the intershell region by artificially adding a partial mixing zone (PMZ) at the deepest extent of each TDU. This method is described by \cite{Lugaro2004} and is similar to that of \cite{GorielyMowlavi2000}.
\cite{Lugaro2012} explore the effect of different PMZ masses in the range $[0,\,0.004]\,\Msun$ on the nucleosynthesis of AGB stars of different initial mass.
The mass of the PMZ is non-zero only in low-mass AGB models ($M\le3\,\Msun$; although as a test the case $M=5.5\,\Msun$ and $\Mpmz=5\times10^{-4}\,\Msun$ is considered). At higher mass (and $Z\le10^{-4}$) the ingestion of protons into the helium-flash-induced convection zone combined with the high temperature at the base of the convective envelope leads to large energy production. This may significantly affect the structure of the star and such an effect cannot be taken into account in the post-processing algorithm (Sect. 2.2 of K10).

\cite{Lugaro2012} distinguish four regimes of neutron-capture process in their models of low-metallicity ($Z=10^{-4}$) AGB stars and each regime dominates in a different mass range. In models of mass above $3\,\Msun$ the $^{22}\Ne$ neutron source dominates and lighter $s$-elements (e.g. strontium) are favoured with respect to heavier $s$-elements (e.g. barium and lead). The $^{13}\C$ neutron source dominates in models of mass below $3\,\Msun$. For masses between $1.75\,\Msun$ and $3\,\Msun$ $\Cth$ burns completely in radiative conditions, neutrons are captured in the thin $\Cth$ layer and consequently the heavier $s$-elements are favoured because of the large number of neutrons per iron seed. On the contrary, for lower masses $\Cth$ burns convectively, neutrons are released over the whole intershell region and thus the neutron-to-iron ratio is lower compared to the case when $\Cth$ burns radiatively. In models with masses below $2.5\,\Msun$, ingestion of protons associated with the first few thermal pulses produce $\Cth$ nuclei, which burn convectively and $s$-elements are produced even in the models without PMZ. This regime dominates for masses below $1.5\,\Msun$.

\cite{Lugaro2012} calculate detailed nucleosynthesis over a grid of 16 initial masses between $0.9$ and 6$\Msun$ for $Z=10^{-4}$. These include abundances predictions for $320$ isotopes from hydrogen through to $^{210}\mathrm{Po}$ as a function of interior mass and time for each single stellar model.  In Sect. \ref{binc:3DUPnuc} we describe the method we used to implement these results in our model of binary population synthesis.

%%%%%%%%%%%%%%%%%%%%%%%%%%%  SECTION 4  %%%%%%%%%%%%%%%%%%%%%%%%%%%
\section{Model of binary evolution and nucleosynthesis}
\label{binc}
To study the chemical compositions of CEMP-$s$ stars, we use the binary evolution and population synthesis code \texttt{binary\_c/nucsyn} described by \cite{Izzard2004,Izzard2006} and recently updated by \cite{Izzard2009, Izzard2010} and \cite{Abate2013}. This code combines a binary-evolution model based on the rapid binary stellar evolution prescriptions of \cite{Hurley2002} with a synthetic nucleosynthesis model developed by \cite{Izzard2004,Izzard2006, Izzard2009}.
In the first part of this section we briefly describe the most important characteristics of our model (Sect. \ref{binc:default}). In the second part we explain how we updated our synthetic nucleosynthesis model with the results of our most recent detailed models of AGB nucleosynthesis (Sect. \ref{binc:3DUPnuc}). In the last part we describe the method used to determine the best fit of the abundances observed in the stars of our sample (Sect. \ref{method}).

\subsection{Input physics}
\label{binc:default}
We describe here the most important input parameters that need to be set in our model of binary evolution and nucleosynthesis. 
In this section we list all the options and in Sect. \ref{results} we specify which assumptions are made in our analysis of each binary system.
\begin{itemize}
\item The wind mass-loss rate up to the AGB phase is parameterised according to the \citet{Reimers75} formula multiplied by a factor $\eta=0.5$ on the first giant branch. During the AGB phase the formula of \cite{VW93} is used, with minimum and maximum values of the wind velocity $v_{\mathrm{w}} = 5 \,\mathrm{km\,s^{-1}}$ and $15 \,\mathrm{km\,s^{-1}}$, respectively. 
\item The variations of angular momentum because of mass loss and the efficiency of the wind mass-transfer process are calculated with two alternative model sets, namely:
	\begin{itemize}
	\item[$\bullet$] 
	{\it Model set A (default):} the angular momentum carried away by the expelled material is computed assuming a spherically symmetric wind \cite[Eq. 4 of][]{Abate2013}. The wind accretion efficiency is calculated according to a model for wind Roche-lobe overflow (WRLOF) that includes the dependence on the mass ratio of the binary system, as in Eq. (9) of \cite{Abate2013}.

	\item[$\bullet$] 
	{\it Model set B:} a model of efficient angular momentum loss is adopted in which the material lost from the binary system carries away a multiple $\gamma=2$ of the average specific orbital angular momentum (\citealp[Eq. 2 of][]{Izzard2010} and \citealp[Eq. 10 of][]{Abate2013}). An extremely efficient wind mass-transfer process is simulated by adopting an enhanced version of the canonical Bondi-Hoyle-Lyttleton prescription for the wind accretion rate, namely Eq. (6) by \cite{BoffinJorissen1988} with $\alpha_{\mathrm{BHL}}=10$, where%
	\footnote{In the paper by \cite{BoffinJorissen1988} $\alpha_{\mathrm{BHL}}$ this constant is indicated simply as $\alpha$} %
	$\alpha_{\mathrm{BHL}}$ is a numerical constant normally between $1$ and $2$.
	\end{itemize}
The purpose of comparing these two model sets is to understand if our models are limited in reproducing the observations by our treatment of the mass-transfer process.
\item Thermohaline mixing is assumed to be efficient: the accreted material mixes instantaneously with the stellar envelope. \cite{Stancliffe2007} suggest that this approximation is typically reasonable. In some cases we relax this assumption and simulate the conditions of highly inefficient thermohaline mixing, in which the accreted material remains on the stellar surface until mixed in by convection.
\item Common-envelope evolution is computed according to the prescription of \cite{Hurley2002} with a free parameter for the common-envelope efficiency of ejection set by default to $\alpha_\mathrm{CE}=1$. We do not include accretion during the common-envelope phase.
\item As initial composition of isotopes up to $^{76}\Ge$ we adopt the abundances predicted by the chemical evolution models of \cite{Kobayashi2011} for solar neighbourhood stars at $[\Fe/\Hy]\approx-2.3$. For heavier isotopes that have not been calculated by \cite{Kobayashi2011}, we adopt the solar distribution of abundances by \cite{Asplund2009} scaled down to metallicity $Z=10^{-4}$.
\begin{figure*}[!t]
	\centering
    \includegraphics[width=\textwidth]{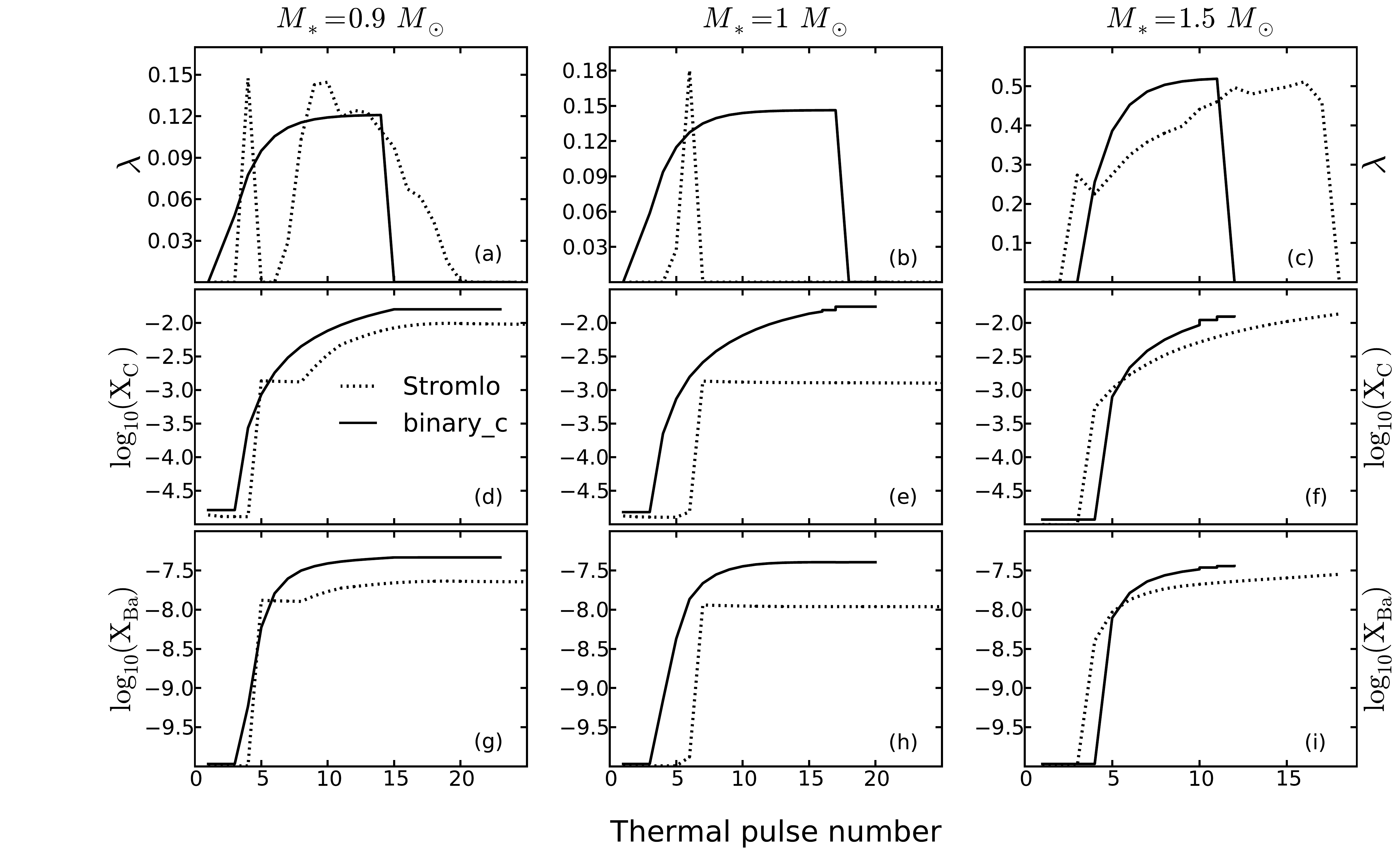}
	\caption{Dredge-up efficiency $\lambda$ (top) and the abundances of carbon and barium (middle and bottom, respectively) computed at each thermal pulse with the $\binc$ and Stromlo codes (solid and dotted lines, respectively) for single star of mass (from left to right) $\Ms=0.9,\, 1,\, 1.5\,\Msun$ and $\Mpmz=2\times10^{-3}\Msun$.}
    \label{fig:lambda-C-Ba-vs-ntp}
\end{figure*}

\item The nucleosynthesis algorithms that compute the evolution of the surface abundances through the first and second dredge-ups are based on the work by \cite{Karakas2002} and \cite{Karakas2007}. A prescription for hot-bottom burning is also included for star more massive than approximately $2.75\,\Msun$. For more details we refer to \cite{Izzard2004,Izzard2006, Izzard2009}. In the next section we describe the treatment of the third dredge-up.
\end{itemize}

\subsection{Third dredge-up nucleosynthesis}
\label{binc:3DUPnuc}

The efficiency of third dredge-up is defined by:
\begin{equation}
\lambda = \frac{M_{\mathrm{DU}}}{\Delta M_{\Hy}}~~, 
\end{equation}
where $M_{\mathrm{DU}}$ is the mass dredged up from the intershell region and $\Delta M_{\Hy}$ is the core-mass growth due to hydrogen burning during the previous interpulse period. Thus, over a whole interpulse period the core grows by:
\begin{equation}
\Delta M_{\mathrm{c}} = \Delta M_{\Hy} - M_{\mathrm{DU}} = (1 - \lambda) \Delta M_{\Hy}~~.
\end{equation}
A TDU episode occurs with efficiency $\lambda$ when $M_{\mathrm{c}}$ exceeds a threshold mass $M_{\mathrm{c,min}}$.
The values of $\lambda$ and $M_{\mathrm{c,min}}$ are functions of mass and metallicity fitted to the detailed models of \cite{Karakas2002}, \cite{Karakas2007} and K10, according to the algorithms explained by \citet[see their Sect. 3.4 for details]{Izzard2004}. 
\cite{Izzard2009} introduced a free parameter in the model, the minimum envelope mass for TDU $\Menvmin$, to study the effects of efficient TDU at masses down to $0.8\,\Msun$. 
To better reproduce $\lambda$ as a function of time from the detailed models of K10 we set,
\[ \frac{\Menvmin}{\Msun} = \left\{
  \begin{array}{l l}
    0.15 & \quad \text{if $\Ms\le1\,\Msun$,}\\
    0.88\,\Ms - 0.73 & \quad \text{if $1\,\Msun<\Ms\le1.25\,\Msun$,}\\
    0.37 & \quad \text{if $\Ms>1.25\,\Msun$.}
  \end{array} \right.\]
With these prescriptions, single stars of initial mass above $0.9\,\Msun$ experience some TDU at metallicity $Z=10^{-4}$, in accordance with the models of K10.

Each TDU modifies the surface abundances of the model AGB star, because material from the stellar interior that has been subject to nuclear processing is mixed into the envelope. For this reason it is essential to know the chemical composition in the intershell region at every thermal pulse if we are to reproduce surface composition of the detailed model with our nucloesynthesis algorithm. 
In our model the abundances in the intershell region of an AGB star of metallicity $Z=10^{-4}$ are stored in a table as a function of three quantities: the mass of the star at the first thermal pulse, the thermal pulse number and a free parameter, $\Mpmz$, that describes the dependence of the chemical composition on the mass of the partial mixing zone according to the Stromlo detailed models. Our table includes all the isotopes provided by the models of \cite{Lugaro2012}.
During the evolution of an AGB star of mass $\Ms$, at each TDU an amount of mass $M_{\mathrm{DU}}$ of the intershell region is instantaneously mixed in the convective envelope. The chemical composition of the dredged-up material is calculated by interpolating values of stellar mass and $\Mpmz$ in our table. As an example of the chemical evolution of the surface of an AGB star due to the TDU episodes, in panels d--i of Fig. \ref{fig:lambda-C-Ba-vs-ntp} we show the abundances of carbon and barium as computed at each thermal pulse with the $\binc$ and Stromlo codes for stars of initial masses $0.9$, $1$ and $1.5\,\Msun$ with $\Mpmz=2\times10^{-3}$. 

\begin{figure}
   \centering
   \includegraphics[width=0.48\textwidth]{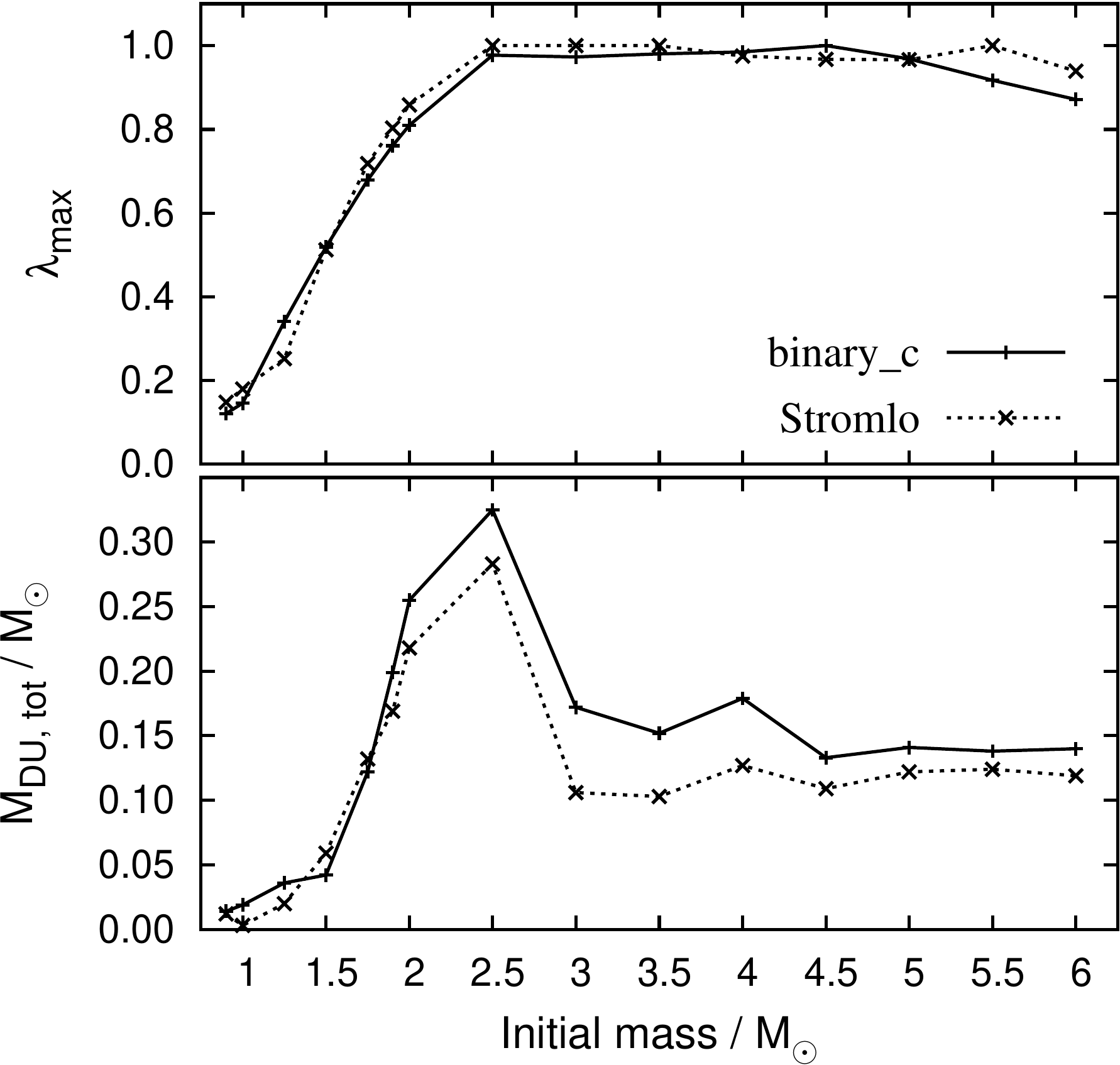}
      \caption{Maximum TDU efficiency $\lambda_{\mathrm{max}}$ (top) and total mass dredged up $M_{\mathrm{DU,\,tot}}$ (bottom) for stellar masses in the range $[0.9,\!6.0]$ as computed with the $\binc$ (solid line) and Stromlo codes (dotted line).}
         \label{fig:lambdamax}
\end{figure}

In Fig. \ref{fig:lambda-C-Ba-vs-ntp}a we show the value of $\lambda$ as a function of the thermal-pulse number computed with our model and with the Stromlo detailed code for a single star of mass $0.9\,\Msun$.
The detailed model predicts no TDU between pulses number 4 and 7 and some TDU up to pulse 19 with a maximum $\lambda \approx 0.14$.
In our model we have increasingly efficient TDU episodes up to pulse 13, when the envelope mass becomes lower than $\Menvmin=0.15$. Despite the difference in $\lambda$ the total amount of mass dredged up in the two models is similar ($0.012\,\Msun$ and $0.014\,\Msun$ with the Stromlo and $\binc$ codes, respectively), as shown in the bottom panel of Fig. \ref{fig:lambdamax}. Consequently, also the surface abundances predicted by two models are similar (Figs. \ref{fig:lambda-C-Ba-vs-ntp}d and \ref{fig:lambda-C-Ba-vs-ntp}g).
Fig. \ref{fig:lambda-C-Ba-vs-ntp}b shows the value of $\lambda$ for a single star of mass $1\,\Msun$. The Stromlo code predicts only two TDUs, a weak one after pulse number 5 and a stronger one after pulse number 6. This peculiar evolution is only observed in the model of a $1\,\Msun$ star, but not for higher masses (see K10 for details), and is probably related to the dependence of the TDU phenomenon on the numerical treatment of convective boundaries, as discussed e.g. by \cite{Frost1996} and \cite{Mowlavi1999}. At mass $M_*=1\,\Msun$ we do not force our model to reproduce $\lambda$ of the detailed model and as a consequence the total mass dredged up with our model is almost $10$ times larger ($0.02\,\Msun$ rather than $0.002\,\Msun$).
Fig. \ref{fig:lambda-C-Ba-vs-ntp}c is the same as Figs. \ref{fig:lambda-C-Ba-vs-ntp}a and \ref{fig:lambda-C-Ba-vs-ntp}b but for a star of mass $1.5\,\Msun$. In this case the main discrepancy is due to the fact that the detailed model predicts more TDU episodes and therefore the total mass dredged up in our model is smaller (bottom panel of Fig. \ref{fig:lambdamax}). However, the final surface abundances of carbon and barium are essentially the same (as shown in Figs. \ref{fig:lambda-C-Ba-vs-ntp}f and \ref{fig:lambda-C-Ba-vs-ntp}i).
At higher masses the two codes are in good agreement, as shown in Fig. \ref{fig:lambdamax} where we plot the maximum value of the TDU efficiency, $\lambda_{\mathrm{max}}$, and the total amount of mass dredged up during the AGB phase, $M_{\mathrm{DU,\,tot}}$, as a function of the initial mass of the star (top and bottom panels, respectively).

\begin{figure*}[!ht]
   \centering
   \includegraphics[width=\textwidth]{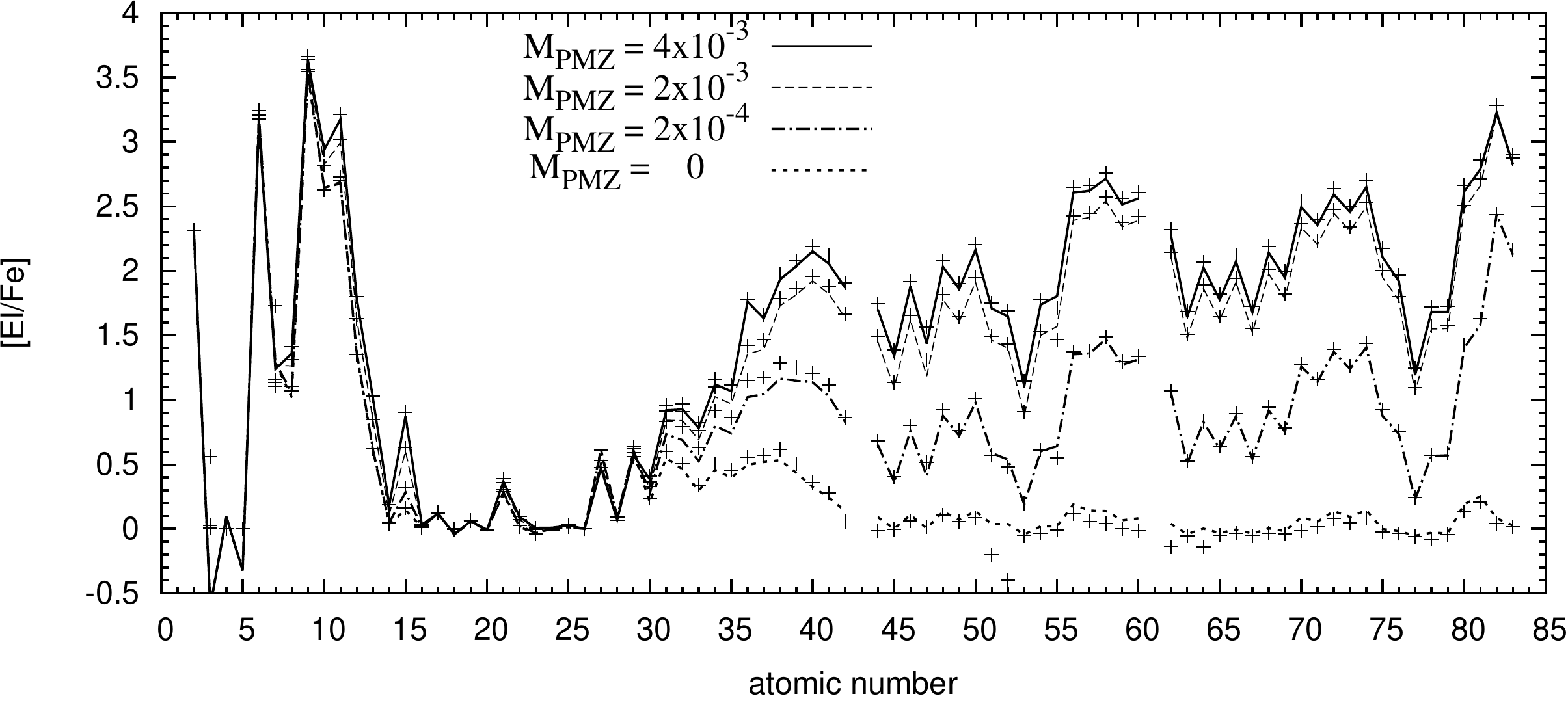}
      \caption{Element abundances relative to iron predicted for a star of mass $\Ms=2\Msun$ with $\Mpmz = 0,\, 2\times10^{-4}, \,2\times10^{-3}, \,4\times10^{-3}\,\Msun$ (respectively: dotted, dot-dashed, dashed, solid line). Plus signs represent the value predicted by the Stromlo model for the same $\Mpmz$.}
         \label{fig:effect_pmz}
\end{figure*}

Because the protons in the PMZ are processed to produce $\Cth$, the parameter $\Mpmz$ plays the same role in the chemical evolution of the AGB star as the efficency of the $\Cth$ pocket discussed by e.g. \cite{Straniero1995}, \cite{Busso1999} and, more recently, \cite{Bisterzo2010}. The main effects of increasing $\Mpmz$ in our model, other conditions being equal, are to bring more $s$-elements to the stellar surface and to produce a distribution of $s$-elements increasingly weighted towards lead.
In Fig. \ref{fig:effect_pmz} we show the element abundances in a $2\,\Msun$ star at the end of the AGB computed with $\binc$ for four different masses of the PMZ. 
While the abundances of light elements differ by about $0.5$ dex at most (Na, Mg, P) and generally remain constant, the abundances of $s$-elements vary significantly between the minimum to the maximum size of the PMZ, $\Mpmz=0\,\Msun$ and $\Mpmz=0.004\,\Msun$, respectively. In Fig. \ref{fig:effect_pmz} we show the results of the Stromlo model for comparison (plus signs): the abundances predicted by the two codes agree to within $0.1$ dex.
We note that below about $2\Msun$ the surface abundances are much less sensitive to $\Mpmz$ and generally only differ by few $0.1$ dex, because for low masses proton-ingestion episodes occur in the detailed models that cause $s$-elements production independently of $\Mpmz$ \cite[we refer to][for details]{Lugaro2012}.

%%

%%%%%%%%%%%%%%%%%%%%%%%%%%%%%%%%%%%%%%%%%%%%%%%
%%

\subsection{Method}
\label{method}

As previously mentioned, the aim of our work is to find for each star in our observed sample a model that reproduces the measured abundances. To this end, we generate a grid of evolution models of $N$ binary stars distributed in the $M_1 - M_2 - \log_{10} a - \Mpmz$ parameter space, where $M_{1,\,2}$ are the initial masses of the primary and of the secondary star, respectively, $a$ is the initial orbital separation of the system and $\Mpmz$ is the mass of the partial mixing zone of any star that undergoes AGB evolution. The grid resolution is equal to 
$N=N_{\mathrm{M1}}\times N_{\mathrm{M2}}\times N_a \times N_{\mathrm{PMZ}}$, 
where we choose $N_{\mathrm{M1}} = 34$, $N_{\mathrm{M2}} = 28$, $N_a = 30$, $N_{\mathrm{PMZ}} = 10$. 
We consider circular orbits in all our models.

The initial parameters are chosen as follows:
\begin{itemize}
\item[$\bullet$] $M_1$ varies in the range [$0.9,6.0$] $\Msun$. Up to $3\,\Msun$ the grid spacing is $\Delta M_1=0.1$ and $\Delta M_1=0.25$ otherwise.
\item[$\bullet$] $M_2$ is uniformly spaced between $0.2\,\Msun$ and $0.9\,\Msun$ ($\Delta M_2=0.025\,\Msun$).
\item[$\bullet$] $a$ varies in the range $[10^2,10^5]$ $\Rsun$. The distribution of separations is uniform in $\log_{10} a$ ($\Delta \log_{10}a/\Rsun=0.1$). In the mass range considered here, stars at wider separation do not interact in our models. All stars in our grid are formed in binary systems.
\item[$\bullet$] $\Mpmz$ is always zero in AGB stars of mass $M\ge3\,\Msun$, otherwise we use the following values: $0$, $10^{-4}$, $2\times10^{-4}$, $5\times10^{-4}$, $6.66\times10^{-4}$, $1\times10^{-3}$, $1.5\times10^{-3}$, $2\times10^{-3}$, $3\times10^{-3}$, $4\times10^{-3}\,\Msun$.
\end{itemize}

We evolve the systems on this grid in parameter space and with different combinations of the physical parameters described in Sect. \ref{binc:default}. 
We select the stars that have not yet become white dwarfs at the evolutionary time $t\ge10$ Gyr and therefore satisfy the surface gravity condition $\loggunits\le5.0$. 
We compare the orbital period, $\logg$ and chemical abundances of the stars that pass these selection criteria with the observations. For each of the $15$ observed stars in our sample the best-fitting model is found with the following iterative procedure.
\begin{enumerate}
\item We first select the modelled binary systems with the right orbital period. Because the observed errors on the periods are generally small (or in some cases this information is not provided in the literature) the working error that we adopt coincides with the spatial resolution in our grid of models, which corresponds approximately to an uncertainty in the orbital period of $\Delta\log_{10}(\Porb/\mathrm{days})=0.15$.
\item Among the modelled stars that belong to binary systems within the correct range of orbital periods, we select those stars that reproduce the observed surface gravity within one $\sigma_g$, where $\sigma_g$ is the observed error on $\logg$. The value of $\logg$ gives an indication of the phase of evolution of a star. 
The constraint on $\logg$ guarantees that we reproduce the evolutionary status of the observed stars.
\item For each of the modelled stars that have passed the selection criteria on $\Porb$ and $\logg$ we calculate the $\chisq$ for the model fit to the abundances with the following equation:
\begin{equation}
\chisq = \sum_i\frac{(A_{i,\mathrm{obs}} - A_{i,\mathrm{mod}})^2}{\sigma_i^2}~~,\label{eq:chi}
\end{equation}
where $A_{i,\mathrm{obs}}$ and $A_{i,\mathrm{mod}}$ are the abundances of element $i$ (defined as in Eq. \ref{eq:abund}) in the observed and modelled star, respectively, and $\sigma_i$ is the observed uncertainty.
We are mostly interested in studying the nucleosynthesis that occurs during the AGB phase, so for this reason we compute $\chisq$ taking into account only the elements that are produced or destroyed by AGB stars. This includes the light elements C, N, O, F, Ne, Na, Mg and all the heavy neutron-capture elements with atomic number in the range $31$ to $82$. In this range are included the peak of light-$s$ elements around atomic number $40$ (strontium, yttrium and zirconium), the peak of heavy-$s$ elements around atomic number $56$ (barium, lanthanum and cerium) and the lead peak. In AGB stars of mass below $\approx3\,\Msun$ nitrogen is produced by the CN cycle when protons from the envelope are mixed into regions of the star where carbon is abundant. The exact amount of nitrogen depends on the extent of this mixing process that is uncertain and is not included in our AGB nucleosynthesis model; however, the total amount of carbon and nitrogen is conserved and therefore when both the elements are observed we fit our models to the combined abundance C+N. In our study we do not consider the elements with atomic number between 13 (aluminium) and 30 (zinc). The abundance of aluminium is not affected by low-mass AGB stars ($M_* \le 3\Msun$), which are the dominant site of the $s$-process, and its production mostly occurs in intermediate-mass AGB stars with hot-bottom burning, $M_* \ge 3\Msun$ \cite[][]{Ventura2009, Karakas2012}. Elements heavier than aluminium up to zinc are not in general produced by AGB stars of any mass range \cite[][]{Karakas2009, Cristallo2011}. The abundances of these elements in very metal-poor stars are expected to be consistent with the chemical composition of the gas cloud from which the stars were formed and to be reproduced by our adopted set of initial abundances, that are based on the results of galactic chemical evolution models at metallicity $Z\approx10^{-4}$. The discrepancies between the observed abundances of elements between aluminium and zinc and our set of initial abundances are discussed more in detail in Paper II.
\item The model that gives the best match to the observed abundances is determined by the minimum value of $\chisq$, $\chimin$. An estimate of how well this model reproduces the observed abundances is given by the reduced $\chisq$, i.e. $\redchi=\chisq/\nu$, where $\nu$ is the number of degrees of freedom. This is calculated as 
\begin{equation}
\nu=N_{\mathrm{obs}} - (n_{\mathrm{fit}}-n_{\mathrm{c}}) = N_{\mathrm{obs}}-3~~, 
\end{equation}
where $N_{\mathrm{obs}}$ is the number of observables, i.e. the number of elements that are used to calculate $\chimin$; $n_{\mathrm{fit}}=5$ is the number of the fitted parameters, i.e. $M_1$, $M_2$, $\Mpmz$, $\logg$, $\Porb$; and $n_{\mathrm{c}}=2$ is the number of observational constraints that are not directly involved in the calculation of $\chisq$ although they limit the range of acceptable models, i.e. $\logg$ and the observed orbital period. Upon visual inspection models with $\chimin/\nu\le3$ appear to fit the observed abundances well, while above this threshold the observations are not reproduced well. The reduced $\chisq$ should not be used for a goodness-of-fit statistical test, because the uncertainties associated with the observed abundances generally only convey the uncertainties of the method adopted to determine the abundances, while systematic errors are not taken into account, which may be caused for example by the errors in the estimates of the temperatures and gravities of the stars (as we mentioned, for example, for star CS22948--027).

\item We determine how well-constrained our best model is by calculating the confidence intervals of its input parameters, $\Mprim$, $\Msec$, $\Pin$ and $\Mpmz$. On our grid of binary models and associated $\chisq$ values, we fix one of the input parameters $p$ and we find the minimum $\chisq$ with respect to the other input parameters. This procedure defines a function that associates a $\chisq$ to each grid value of the parameter $p$. The minimum of this function is equal to the $\chisq$ of the best fit, $\chimin$. A confidence region is defined as an interval of $p$ within which the difference $\chisq-\chimin$ is below a certain threshold. 
If the measurement errors are Gaussian and our model reproduces the data, then the probability distribution of $\chisq$ is also Gaussian. In this case, the threshold $\Delta\chisq=1$ corresponds to the confidence interval of $68.3\%$ probability that the actual $p$ is in this interval. Similarly, the thresholds $\Delta\chisq=4$ and $\Delta\chisq=9$ correspond to the confidence intervals of $95.4\%$ and $99.7\%$ probability, respectively. However, because systematic errors are generally not taken into account in the estimated observational errors, the thresholds $\Delta\chisq=1,\,4,\,9$ should not be used to calculate the theoretical Gaussian probabilities. In the analysis of our sample stars we note that models with $\chisq$ below the threshold $\Delta\chisq = 4$ have surface abundances that are hard to distinguish by eye from the best-fit values, whereas higher $\chisq$ are found for models clearly distinct and worse than the best model. Therefore, to determine the confidence intervals of our input parameters we adopt $\Delta\chisq = 4$. In case for one parameter no model other than the best fit satisfies the condition $\Delta\chisq<4$, the confidence interval is assumed to be half of the grid resolution for that parameter.
\end{enumerate}

In points $1-4$ of our procedure we do not include any constraint on $\Teff$, which has a strong dependence on the metallicity. The observed metallicities of the stars in our sample vary by up to a factor of five. On the other hand, in our model we keep the metallicity constant because the nucleosynthesis is valid for $Z=10^{-4}$ and hence we do not expect to be able to reproduce~$\Teff$.

In our study we do not take into account that most neutron-capture elements are not purely produced by the $s$-process but also have an $r$-process component \cite[][]{Arlandini1999, Bisterzo2011}. The origin of the $r$-process elements in metal-poor stars is still unclear \citep{Sneden1994, Jonsell2006, Sneden2008, Lugaro2009, Masseron2010} and a thorough analysis of this issue is beyond the scope of this paper.

%%%%%%%%%%%%%%%%%%%%%%%%%%%  SECTION 5  %%%%%%%%%%%%%%%%%%%%%%%%%%%
\section{Results}
\label{results}
%
%%%%%%%%%%%%%%%%%%%%% TABLES %%%%%%%%%%%%%%%%%%%%%%%%
\begin{table*}
\caption{Physical parameters of the modelled stars computed with model set A with WRLOF wind-accretion rate and spherically symmetric wind.}
\label{tab:bestfit_WRLOFq}
\centering
\begin{tabular}{ l | c c c c c | c c c c | c c c }
\hline
\hspace{9.5mm}ID & $\Mprim$ & $\Msec$ & $\Mpmz$ & $\Pin$ & $\aCE$ & $\Pf$ & $\log_{10}g$ & $\Teff$ & $\Delta \Macc$ & $\chimin$ & $\nu$ & $\chimin/\nu$ \\
\hline
BD$+04^{\circ}2466$ & $1.1$ & $0.76$ & $2\times10^{-3}$ & $3.19\times10^3$ & & $4.28\times10^3$ & $1.61$ & $4600$ & $0.09$ & $13.4$ & $8$ & $1.7$ \\
CS22942--019 & $1.4$ & $0.76$ & $4\times10^{-3}$ & $2.10\times10^3$ & & $2.87\times10^3$ & $2.60$ & $5000$ & $0.10$ & $74.3$ & $10$ & $7.4$ \\
CS22948--027 & $0.9$ & $0.81$ & $2\times10^{-3}$ & $4.19\times10^2$ & & $4.76\times10^2$ & $2.20$ & $4900$ & $0.12$ & $82.8$ & $10$ & $8.3$ \\
CS22956--028 & $0.9$ & $0.84$ & $2\times10^{-3}$ & $1.17\times10^3$ & & $1.37\times10^3$ & $3.82$ & $6900$ & $0.10$*& $50.9$ & $ 2$ & $25.4$ \\
CS22964--161A& $1.6$ & $0.79$ & $2\times10^{-3}$ &$2.74\times10^5$&  & $5.60\times10^5$ &$3.50$&$5800$&$0.04$&$19.5$& $ 8$ & $2.5$ \\	
CS22964--161B& $1.6$ & $0.71$ & $2\times10^{-3}$ &$2.74\times10^5$&  & $5.60\times10^5$ &$4.48$&$6200$&$0.02$&$11.8$& $9$ & $1.3$ \\
CS29497--030 & $0.9$ & $0.84$ & $2\times10^{-3}$ & $3.37\times10^2$& & $4.04\times10^2$ & $3.63$ & $6500$ & $0.14$*& $145.6$ & $17$ & $8.6$ \\
CS29497--034 & $1.9$ & $0.76$ & $2\times10^{-3}$ & $2.67\times10^3$ & & $3.35\times10^3$ & $1.80$ & $4700$ & $0.11$ & $117.6$ & $10$ & $11.8$ \\
CS29509--027 & $2.9$ & $0.71$ & $2\times10^{-3}$ & $3.09\times10^3$ & $1.0$ & $1.65\times10^2$ & $4.14$ & $7000$ & $0.08$& $ 4.6$ & $ 1$ & $4.6$ \\
HD$198269$    & $0.9$ & $0.79$ & $3\times10^{-3}$ & $1.19\times10^3$ & & $1.43\times10^3$ & $1.55$ & $4600$ & $0.06$& $32.4$ & $ 6$ & $5.4$ \\
HD$201626$    & $0.9$ & $0.81$ & $2\times10^{-3}$ & $4.19\times10^2$ & & $4.75\times10^2$ & $2.23$ & $4900$ & $0.12$& $30.1$ & $ 6$ & $5.0$ \\
HD$224959$    & $0.9$ & $0.79$ & $3\times10^{-3}$ & $1.19\times10^3$ & & $1.42\times10^3$ & $2.05$ & $4800$ & $0.06$& $340.1$ & $9$ & $37.8$ \\
HE0024--2523 & $1.9$ & $0.71$ & $2\times10^{-3}$ & $2.57\times10^3$ & $0.07$& $3.33$ & $4.40$ & $6500$ & $0.06$& $203.6$& $ 5$ & $40.7$ \\
HE0507--1430 & $0.9$ & $0.81$ & $0             $ & $4.19\times10^2$ & & $5.16\times10^2$ & $0.70$ & $4300$ & $0.12$& $26.5$& $ 0$ & $-$ \\
LP625--44 	  & $1.9$ & $0.76$ & $4\times10^{-3}$ & $2.67\times10^3$ & & $4.03\times10^3$ & $2.95$ & $5100$ & $0.11$& $374.0$& $18$ & $20.8$ \\  
\hline
\end{tabular}
\end{table*}

\begin{table*}
\caption{Physical parameters of the modelled stars computed with model set B in which we adopt an efficient Bondi-Hoyle-Lyttleton wind-accretion rate ($\alpha_{\mathrm{BHL}}=10$) and efficient angular-momentum loss.}
\label{tab:bestfit_BHL}
\centering
\begin{tabular}{ l | c c c c c | c c c c | c c c }
\hline
\hline
\hspace{9.5mm}ID & $\Mprim$ & $\Msec$ & $\Mpmz$ & $\Pin$ & $\aCE$ & $\Pf$ & $\log_{10}g$ & $\Teff$ & $\Delta \Macc$ & $\chimin$ & $\nu$ & $\chimin/\nu$ \\
\hline
BD$+04^{\circ}2466$ & $0.9$ & $0.79$ & $3\times10^{-3}$ & $4.74\times10^3$&  & $4.25\times10^3$ & $1.70$ & $4700$ & $0.05$ & $25.7$ & $8$ & $3.2$ \\
CS22942--019 & $1.8$ & $0.54$ & $2\times10^{-4}$ & $1.67\times10^4$ & & $3.23\times10^3$ & $2.26$ & $4900$ & $0.29$ & $21.0$ & $10$ & $2.1$ \\
CS22948--027 & $1.5$ & $0.61$ & $1\times10^{-3}$ & $6.72\times10^3$ & $1.0$ & $3.69\times10^2$ & $2.20$ & $4900$ & $0.27$ & $31.9$ & $10$ & $3.2$ \\
CS22956--028 & $0.9$ & $0.84$ & $2\times10^{-3}$ & $1.17\times10^3$ & & $1.05\times10^3$ & $3.90$ & $7100$ & $0.10$*& $56.3$ & $ 2$ & $28.2$ \\
CS22964--161A& $1.6$ & $0.79$ & $2\times10^{-3}$ &$1.30\times10^5$ & & $1.56\times10^5$ &$3.50$&$5800$&$0.04$&$19.6$& $ 8$ & $2.5$ \\  
CS22964--161B& $1.6$ & $0.71$ & $2\times10^{-3}$ &$1.30\times10^5$ & & $1.56\times10^5$ &$4.48$&$6200$&$0.02$&$11.8$& $10$ & $1.3$ \\
CS29497--030 & $1.5$ & $0.56$ & $2\times10^{-3}$ & $7.17\times10^3$ & $1.0$ & $3.00\times10^2$ & $4.34$ & $6700$ & $0.22$*& $116.0$ & $17$ & $6.8$\\
CS29497--034 & $1.6$ & $0.56$ & $2\times10^{-3}$ & $1.18\times10^4$ & & $3.35\times10^3$ & $1.77$ & $4700$ & $0.31$ & $34.1$ & $10$ & $3.4$ \\
CS29509--027 & $2.9$ & $0.59$ & $5\times10^{-4}$ & $2.62\times10^4$ & $1.0$ & $1.67\times10^2$ & $4.28$ & $6900$ & $0.19$& $ 5.0$ & $ 1$ & $5.0$ \\
HD$198269$    & $1.0$ & $0.81$ & $3\times10^{-3}$ & $1.62\times10^3$ & & $1.17\times10^3$ & $1.05$ & $4400$ & $0.13$& $18.0$ & $ 6$ & $3.0$ \\
HD$201626$    & $2.6$ & $0.64$ & $4\times10^{-3}$ & $2.72\times10^4$ &$1.0$& $3.55\times10^2$ & $2.50$ & $5000$ & $0.22$& $13.6$ & $ 6$ & $2.3$ \\
HD$224959$    & $1.2$ & $0.61$ & $4\times10^{-3}$ & $3.24\times10^3$ & & $1.41\times10^3$ & $2.12$ & $4800$ & $0.34$& $72.2$ & $9$ & $8.0$\\
HE0024--2523 & $1.1$ & $0.69$ & $0             $ & $2.15\times10^3$ & $0.03$& $3.48$ & $4.40$ & $6600$ & $0.11$& $50.0$& $ 5$ & $10.0$ \\
HE0507--1430 & $1.8$ & $0.61$ & $0             $ & $1.12\times10^4$ & $1.0$& $3.58\times10^2$ & $0.70$ & $4300$ & $0.28$& $7.6$& $ 0$ & $-$ \\
LP625--44 	  & $1.8$ & $0.56$ & $4\times10^{-3}$ & $1.66\times10^4$ & & $3.65\times10^3$ & $2.94$ & $5100$ & $0.30$& $194.4$& $18$ & $10.8$ \\  
\hline
\end{tabular}
\tablefoot{All the masses are expressed in units of $\Msun$, periods are in days, $\Teff$ in K, and $g$ is in units of cm s$^{-2}$; $\nu$ is the number of degree of freedom of the fit; $\aCE$ is shown for modelled systems that experience a common-envelope phase.\\
(*) inefficient thermohaline mixing, the accreted material stays on the stellar surface.\\
}
\end{table*}

\begin{table*}[!ht]
\caption{Confidence intervals of the input parameters of our model stars with $\chisq/\nu\le3$.}
\label{tab:confi}
\centering
\begin{tabular}{ l  c c | c c c | c c c | c c c | c c c  }
\hline
\hline
\hspace{9.5mm}ID & model set & $\chimin/\nu$ & \multicolumn{3}{c|}{$\Mprim/\Msun$} &  \multicolumn{3}{c|}{$\Msec/\Msun$} & \multicolumn{3}{c|}{$\Mpmz/(10^{-3}\Msun)$} & \multicolumn{3}{c}{$\Pin/(10^3$days)} \\
 & & & best & min & max & best & min & max & best & min & max & best & min & max \\
\hline
BD$+04^{\circ}2466$ & A & $1.7$ & $1.10$ & $0.95$ & $1.15$ & $0.76$ & $0.73$ & $0.85$ & $2.0$ & $0.05$ & $4.0$ & $3.19$ & $2.32$ & $3.92$ \\
 &  &  &  & $1.45$ & $1.65$ & & & & & & & & & \\
CS22942-019 & B & $2.1$ & $1.80$ & $1.75$ & $1.85$ & $0.54$ & $0.53$ & $0.58$ & $0.2$ & $0.0$ & $0.4$ & $16.7$ & $14.1$ & $19.8$ \\
CS22964-161A & B & $2.5$ & $1.60$ & $0.95$ & $1.65$ & $0.79$ & $0.75$ & $0.85$ & $2.0$ & $0.58$ & $4.0$ & $130$ & $103$ & $232$ \\
CS22964-161B & B & $1.3$ & $1.60$ & $0.95$ & $1.65$ & $0.71$ & $0.68$ & $0.85$ & $2.0$ & $0.58$ & $4.0$ & $130$ & $103$ & $232$ \\
HD$198269$ & B & $3.0$ & $1.00$ & $0.95$ & $1.05$ & $0.81$ & $0.73$ & $0.83$ & $3.0$ & $0.80$ & $4.0$ & $1.62$ & $1.36$ & $1.93$ \\
HD$201626$ & B & $2.3$ & $2.60$ & $1.95$ & $2.95$ & $0.64$ & $0.58$ & $0.65$ & $4.0$ & $0.80$ & $4.0$ & $27.2$ & $15.2$ & $36.3$ \\
 &  &  &  & $1.55$ & $1.65$ &  &  &  &  &  &  &  & $6.98$ & $9.86$ \\
\hline
\end{tabular}
\end{table*}
%%%%%%%%%%%%%%%%%%%%%%%%%%%%%%%%%%%%%%%%%%%%%%%%%%%%%

We determine the best-fitting models to the observed abundances of the $15$ stars in our sample. In Tables \ref{tab:bestfit_WRLOFq} and \ref{tab:bestfit_BHL} we summarise the results obtained with model sets A and B, respectively, as follows. 

\begin{itemize}
\item[$\bullet$] Columns $2-6$: the initial masses of the modelled stars, the mass of the PMZ, the initial orbital period and, for binary systems that undergo common-envelope evolution, the adopted value of the common-envelope efficiency.
\item[$\bullet$] Columns $7-12$: the results of the fit, namely the period of the modelled binary, $\Pf$, at the moment when the modelled secondary star best reproduces $\logg$ and the observed abundances, the surface gravity, the effective temperature, the amount of mass accreted by the secondary $\Delta \Macc$, the minimum value of $\chisq$, $\chimin$, the number of degrees of freedom of the fit, $\nu$, and the reduced $\chisq$, $\chimin/\nu$. 
\end{itemize}

In the next section we describe the main results of our study, while in Sect. \ref{subsec:CS22942-019} and \ref{subsec:CS29497-030} we discuss in detail the models of the $r$-normal CEMP-$s$ star CS22942--019 and the CEMP-$s/r$ star CS29497--030 as an example of our analysis. In Sect. \ref{subsec:CS22964-161AB} we discuss the method adopted to model the CEMP-$s$ binary star CS22964-161A,B. The plots of the best-fitting models of the remaining $11$ stars of our sample are shown in Appendix \ref{app:A}.

%%%%%%%%%%%%%%%%%%%%%%%%%
\subsection{Main results}
The observed orbital periods provide strong constraints to the initial orbital periods in our models and consequently to the initial primary masses. 
In our model set A (with a spherically symmetric wind), the systems typically widen in response to mass loss. Hence, the final orbital periods reproduce the observations only if the initial periods are relatively short (less than a few thousand days). Consequently, the initial primary masses are relatively low (generally $\Mprim\le1.1\Msun$), because the primary stars have to evolve up to the AGB without overfilling their Roche lobes, otherwise the binary systems start a common-envelope phase without mass accretion.
On the other hand, with model set B (efficient angular momentum loss) the systems shrink in response to mass loss, therefore longer initial orbital periods are possible and consequently higher primary masses. 

Six stars in our sample have periods shorter than $500$ days. For these systems our model set B predicts that initially the periods are up to a hundred times longer. During the AGB phase the primary star loses about $60-70\%$ of its envelope mass in the wind before filling its Roche lobe. The envelope mass that is left forms a common envelope, the ejection of which requires the system to lose angular momentum and shrink to the observed orbital period. In star HD224959 (Fig. \ref{fig:HD224959}) the common-envelope phase is preceded by a short phase of stable Roche-lobe overflow, during which $0.1\Msun$ are accreted by the secondary star. Star HE0024--2523 (Fig. \ref{fig:HE0024-2523}) has an observed orbital period of $3.14$ days. In this close orbit all primary stars of our grid overfill the Roche lobe while ascending the red giant branch. An initially wider orbit is necessary for the primary star to evolve to the AGB phase. During this phase some material is transferred onto the secondary star before the system enters in a common envelope that is ejected with very low efficiency ($\aCE=0.07$ or $0.03$ for model sets A and B, respectively). An alternative interpretation that does not require such a low common-envelope efficiency is discussed in Sect. \ref{disc_binev}.

Eight stars in our sample have passed the main-sequence turnoff and have low surface gravity. As a consequence, large amounts of accreted material are required to reduce the dilution caused by the first dredge-up and reproduce the enriched abundances of carbon and $s$-elements. At periods shorter than about $5,\!000$ days small amounts of mass are accreted with our model set A, while in model set B (with enhanced BHL accretion efficiency) the accreted mass is up to a few tenths of a solar mass. As a consequence of the constraints on the period, $\Mprim$ and $\Delta\Macc$, the best-fitting models to the abundances of ten stars are found with model set B. Model set A predicts a significantly better fit than set B only for star BD$+04^{\circ}2466$, which is the longest-period binary in our sample and exhibits relatively small enhancements ($[\El/\Fe]<2$ for all observed elements, Fig. \ref{fig:BD+04o2466}). Hence, a relatively small accreted mass ($\Delta\Macc<0.1\Msun$) is sufficient to reproduce the observed abundances.

We obtain $\chimin/\nu\le3$ only for six stars. The confidence intervals of the input parameters of these systems are summarised in Table \ref{tab:confi}. Multiple local minima occur in the $\chisq$ distributions of the primary masses and initial separation of stars BD$04^{\circ}2466$ and HD$201626$. These minima are found because different combinations of initial parameters result in model stars with similar surface abundances and hence similar $\chisq$.  A few examples of the confidence intervals determined in our study are discussed in Sect. \ref{subsec:CS22942-019} and in Appendix~\ref{app:B}.

%%%%%%%%%%%%%%%%%%%%%%%%%%%%%%%
\subsection{Example 1: the $r$-normal CEMP-$s$ star CS22942--019}
\label{subsec:CS22942-019}

%%%
\begin{figure*}[!t]
\begin{center}
\includegraphics[angle=0, width=0.98\textwidth]{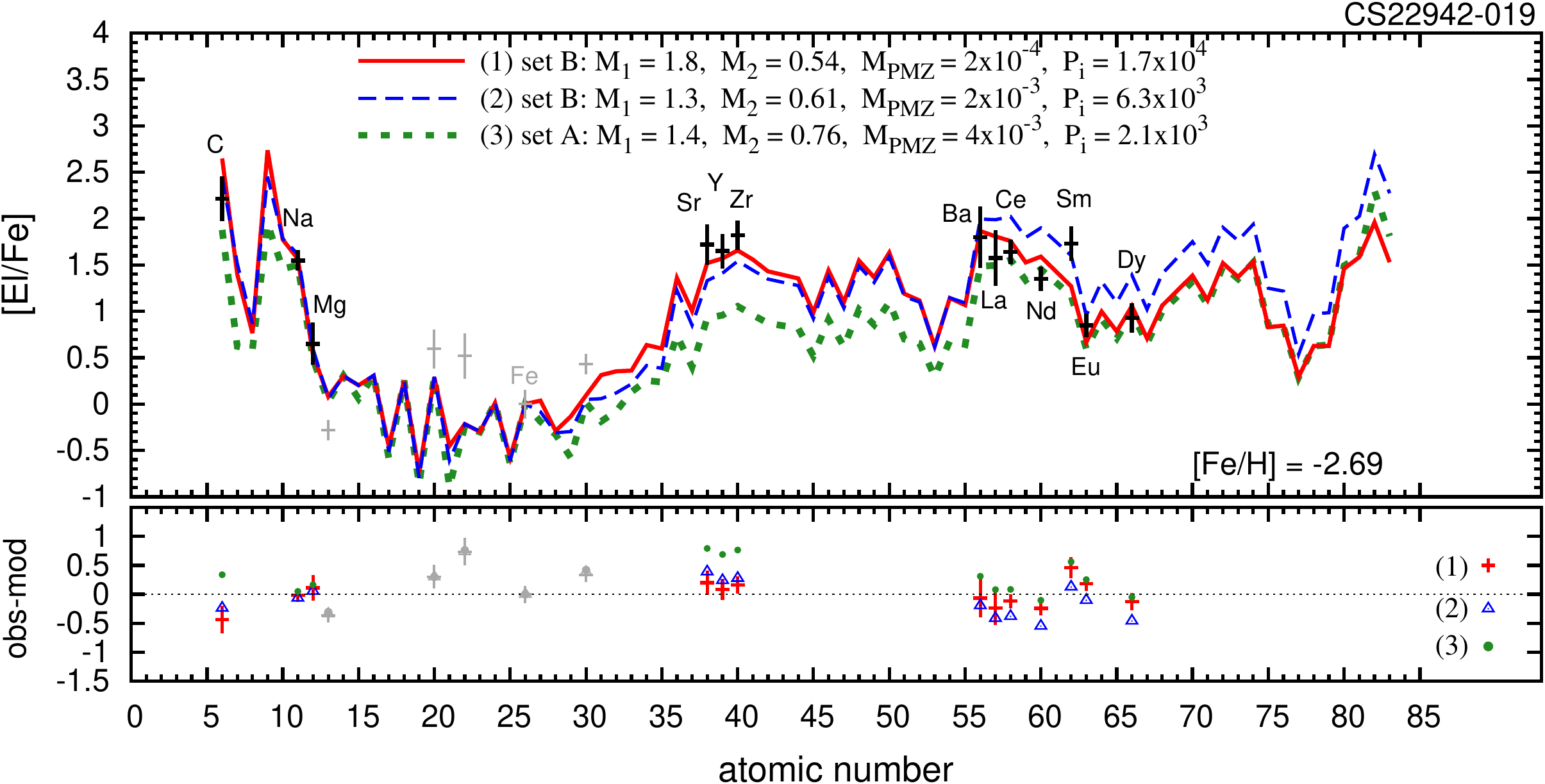}
\end{center}
\caption{Best fitting model to star CS22942--019. {\it Points with error bars}: observed abundances, in black the elements used to determine the best fit and in grey the other elements. {\it Red solid line}: the best-fitting model, found with model set B and $\Mprim=1.8\,\Msun$, $\Msec=0.54\,\Msun$ and $\Mpmz=2\times10^{-4}\,\Msun$. {\it Blue dashed line}: alternative fit with model set B and $\Mprim=1.3\,\Msun$, $\Msec=0.61\,\Msun$ and $\Mpmz=2\times10^{-3}\Msun$ (see text). {\it Green dotted line}: best fit adopting model set A.
{\it Lower panel}: the residuals of the three models are shown as red plus signs with error bars, blue triangles, and green dots, respectively.}
\label{fig:CS22942-019}
\end{figure*}

%%%
\begin{figure}[!t]
\centering
\includegraphics[angle=0, width=0.48\textwidth]{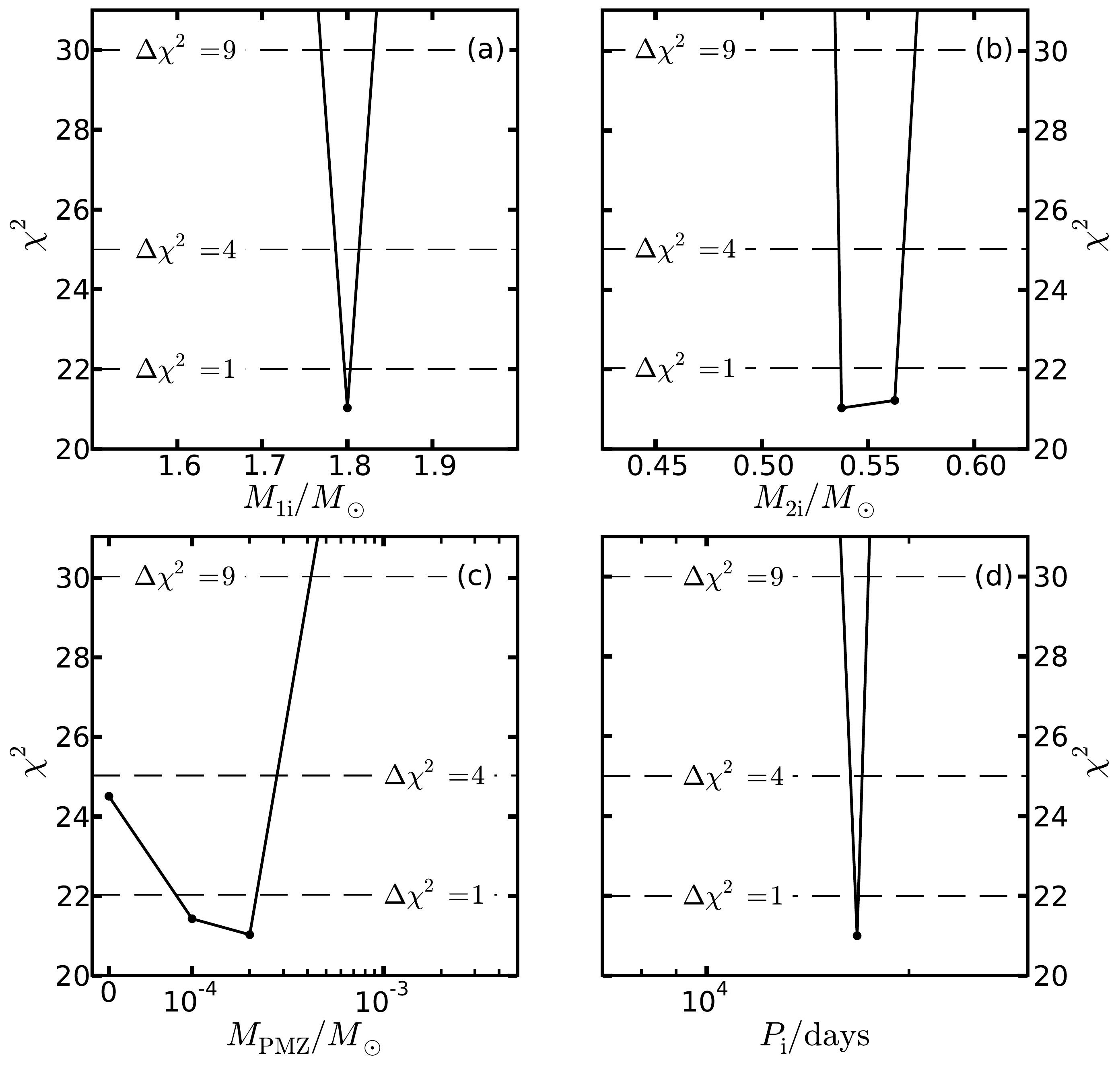}
\caption{One-dimensional confidence intervals of the input parameters for model star CS22942--019. Panels a--d show the initial primary and secondary masses, $\Mprim$ and $\Msec$, the PMZ mass, $\Mpmz$, and the initial period, $\Pin$, respectively. Long-dashed lines indicate the thresholds $\Delta\chisq=1,\,4,\,9$.}
\label{fig:confiCS22942-019}
\end{figure}
In Fig. \ref{fig:CS22942-019} we show the observed and modelled abundances of star CS22942--019. Black points with solid error bars show the abundances of the elements taken into account in Eq. (\ref{eq:chi}) to determine the best fit. The other observed elements are shown in grey. Our best model (red solid line) is found with model set B, in which a primary star of mass $\Mprim=1.8\,\Msun$ transfers material to a $0.54\,\Msun$ companion (Table \ref{tab:bestfit_BHL}). In the lower panel of Fig. \ref{fig:CS22942-019} we show the residuals calculated as the difference between the observed and the modelled abundance of each element. Our best model is found combining three ingredients: an initially wide separation, so that the primary star does not fill its Roche lobe, high mass accretion rate, and efficient angular momentum loss, which is required to shrink the initially wide system to the observed period.

In our best model computed with our default model set A (green dotted line in Fig. \ref{fig:CS22942-019}) the binary system widens instead of shrinking as in model set B. Consequently, the initial orbital separation needs to be shorter ($\Pin\approx2100$ days) to avoid filling the Roche lobe and therefore a lower $\Mprim$ is necessary ($\Mprim=1.2-1.4\,\Msun$). With these initial parameters our WRLOF prescription is not very efficient: the secondary star accretes a smaller amount of material compared to our best model, approximately $\Delta\Macc=0.1\,\Msun$, and therefore it needs to be initially more massive, $\Msec=0.76\,\Msun$, to reproduce the observed $\log g$. The accreted material is more diluted in the secondary star and we underestimate the observed abundance of carbon, the light-$s$ elements and barium even when assuming the largest PMZ available in our model, $\Mpmz=4\times10^{-3}$. Consequently the fit is much poorer, $\chimin=74$.

The choice of a relatively massive primary star in model set B ($\Mprim=1.8\,\Msun$) implies that the abundances of almost all the elements are reproduced with $\Mpmz=2\times10^{-4}$. A primary star more massive than $1.8\,\Msun$ overproduces sodium (even with $\Mpmz=0$) and typically produces excessively abundant heavy-$s$ elements, in contrast with the observations, which give $[\hs/\ls]\approx 0$. \cite{Lugaro2012} consider $\Mpmz=2\times10^{-3}\,\Msun$ as the standard PMZ mass and  most models in Tables \ref{tab:bestfit_WRLOFq} and \ref{tab:bestfit_BHL} have a relatively large PMZ (typically few $10^{-3}\,\Msun$). In Fig. \ref{fig:CS22942-019} we compare our best fit with the best model that adopts $\Mpmz=2\times10^{-3}\,\Msun$ (blue dashed line). The initial parameters of this model are $\Mprim=1.3\,\Msun$, $\Msec= 0.61\,\Msun$, $\Pin=6.28\times10^3$ days, and the other assumptions are the same as in the best model. With this model the light-$s$ elements are underestimated while most of the heavy-$s$ elements are overestimated 
and as a result $\chisq=54$.

In Fig. \ref{fig:confiCS22942-019} we show the confidence intervals of the input parameters of model star CS22942--019 computed with set B. Panels a--d show the one-dimensional confidence intervals of the initial primary and secondary masses, PMZ mass and initial orbital period, respectively. No model is found with $\Delta\chisq<9$ if $\Mprim$ or $\Pin$ differ from the best-fitting values and few models fulfil the condition $\Delta\chisq<4$ for a variation of $\Msec$ or $\Mpmz$. These results do not imply that the initial parameters of the progenitor system of CS22942--019 are determined with small uncertainty. The narrow confidence intervals that we derive (Table \ref{tab:confi}) are a consequence of the constraint put by the observed period. With a $1.8\Msun$ primary star we find only one initial period in our grid of models for which a binary system evolves into a system with the observed orbital period. If a different initial primary mass is selected, no combination of the other three parameters is found that reproduces the measured period and the observed abundances with $\Delta\chisq<4$. In Paper~II we show that the confidence intervals of $\Mpmz$, $\Pin$ and $\Msec$ are larger if we release the period constraint.

%%%%%%%%%%%%%%%%%%%%%%%%%%%%%%%
\subsection{Example 2: the CEMP-$s/r$ star CS29497--030}
\label{subsec:CS29497-030}

%%%
\begin{figure*}[!t]
\begin{center}
\includegraphics[angle=0, width=0.98\textwidth]{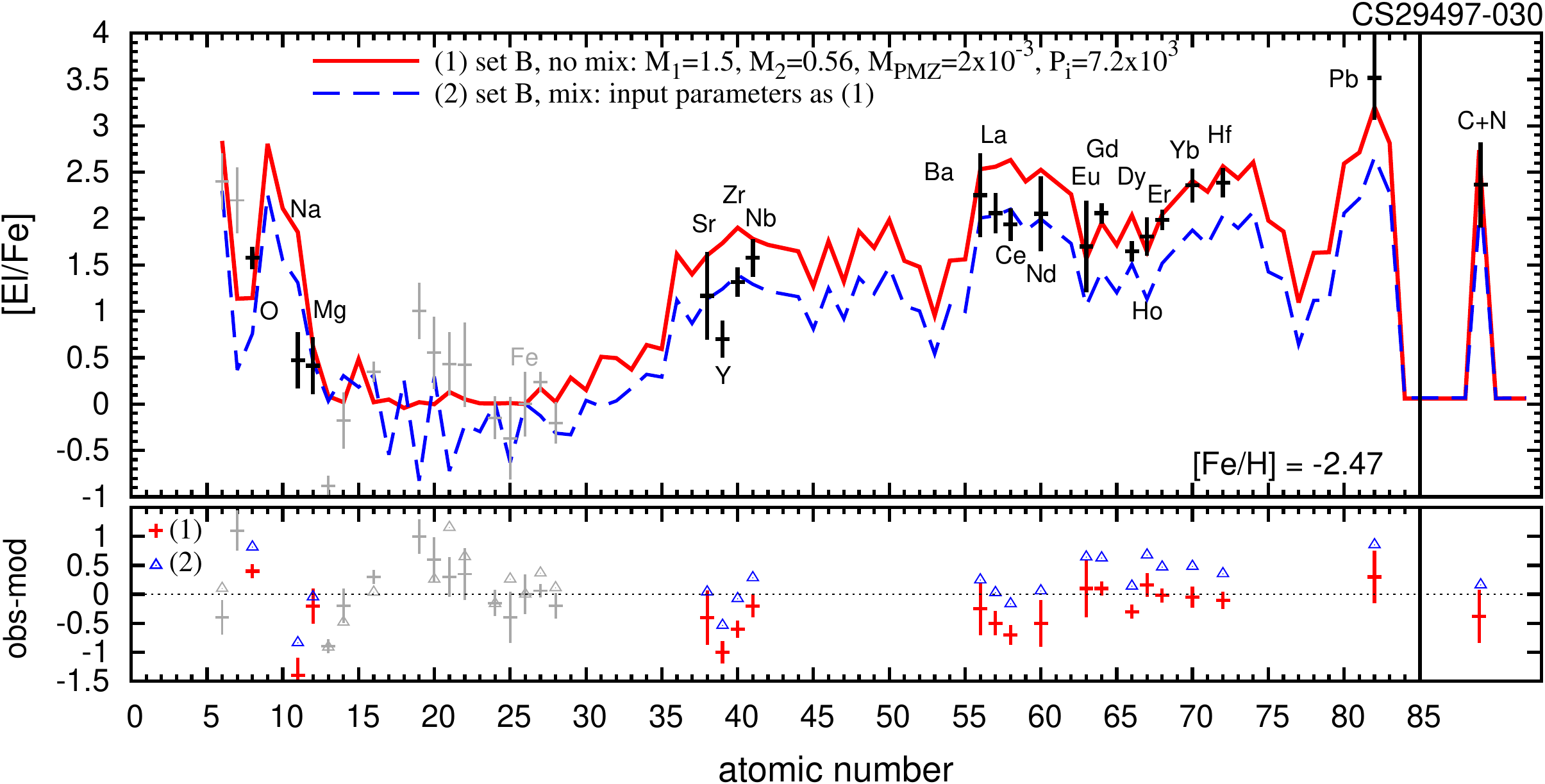}
\end{center}
\caption{{\it Points with error bars}: as Fig. \ref{fig:CS22942-019} for star CS29497--034. {\it Red solid line}: best-fitting model found with model set B and inefficient mixing of the material accreted by the secondary star. {\it Blue dashed line}: same input parameters as the best model but assuming efficient thermohaline mixing. In the right panel the combined abundance of carbon and nitrogen is shown.}
\label{fig:CS29497-030}
\end{figure*}

The observed period of star CS29497--030 in Fig. \ref{fig:CS29497-030} is rather short, $\Porb=342$ days. In our model set B the binary system is initially wider (about $7200$ days). The primary star ($\Mprim=1.5\Msun$) ascends the AGB and loses approximately $0.53\Msun$ in the wind, which is partly accreted by the secondary star ($\Delta\Macc=0.22\Msun$). Subsequently, when its residual envelope mass is about $0.25\Msun$, the primary star fills its Roche lobe and the system enters a common-envelope phase. The ejection of the common envelope, modelled with $\aCE=1$, shrinks the system to the period $\Pf=300$ days that reproduces the observed value within the uncertainty of our grid. The common-envelope phase is required because the wind of the primary star does not carry away enough angular momentum even with our model set B.

With our model set A we find a binary system with almost equal initial primary and secondary masses ($\Mprim=0.9\Msun\,$, $\Msec=0.84\Msun$) and short initial period ($\Pin=337$ days) that expands as an effect of mass loss and does not undergo a common-envelope phase. The amount of mass transferred is small compared to model set B ($\Delta\Macc=0.14\Msun$) and the fit is significantly worse ($\chimin\approx146$, whereas $\chimin=116$ for model~set~B).

Star CS29497--030 is strongly enhanced in carbon, heavy-$s$ elements, $r$-process elements and lead. On the other hand, sodium, magnesium and the light-$s$ elements are only weakly or mildly enhanced (between $0.4$ and $1.5$ dex). Similar abundance distributions are also observed in the other CEMP-$s/r$ stars of our sample (CS22948--027, CS29497--034, HD224959 and LP625--44). Our best model (red solid line in Fig. \ref{fig:CS29497-030}) reproduces the abundances of barium, most $r$-elements and lead within the observational uncertainty but it overestimates the abundance of sodium (by approximately $1.5$ dex), light-$s$ elements, lanthanum and cerium (by $0.3-1$ dex). Our best model is computed assuming that non-convective mixing mechanisms of the accreted material, such as thermohaline mixing, are inefficient. Consequently, the transferred material remains on the surface of the secondary star because the relatively high observed gravity, $\loggunits=4.0$, indicates that CS29497--030 is still on the main sequence and therefore it has not yet undergone the first dredge-up. 

The blue dashed line in Fig. \ref{fig:CS29497-030} shows the abundances predicted by a model with the same input parameters as in the best model and efficient thermohaline mixing. This model reproduces the abundances of C+N, Mg, light-$s$ elements, La, Ce and Nd better than the best-fitting model, but the $\chisq$ is higher ($\chisq=167$) because it fails to reproduce the eight elements between europium and lead. Except for lead, which is off by less then $2\sigma$, these elements are mostly produced by the $r$-process. If we exclude the $r$-elements from the calculation of the fit we find the same initial parameters as in the best model. The reduced $\chisq$ is still high ($\chisq/\nu=7.4$) because it is dominated by the discrepancy in the abundance of sodium, which is overestimated by approximately $1$ dex. Because both models in Fig. \ref{fig:CS29497-030} do not reproduce the observed abundances well ($\chimin/\nu>3$) we do not have strong constraints on the efficiency of non-convective mixing processes in CS29497--030.
%

%%%%%%%%%%%%%%%%%%%%%%%%%%%%%%%%%%%%%%%%%%%%%%%%%%%%%%%%%%%%%%%%
\subsection{Example 3: the CEMP-$s$ binary star CS22964--161A,B}
\label{subsec:CS22964-161AB}
%

%%%%%%%%%%%%%%%%%%%%%%%%%%%%%%%
\begin{figure*}[!t]
\centering
\includegraphics[angle=0, width=0.48\textwidth]{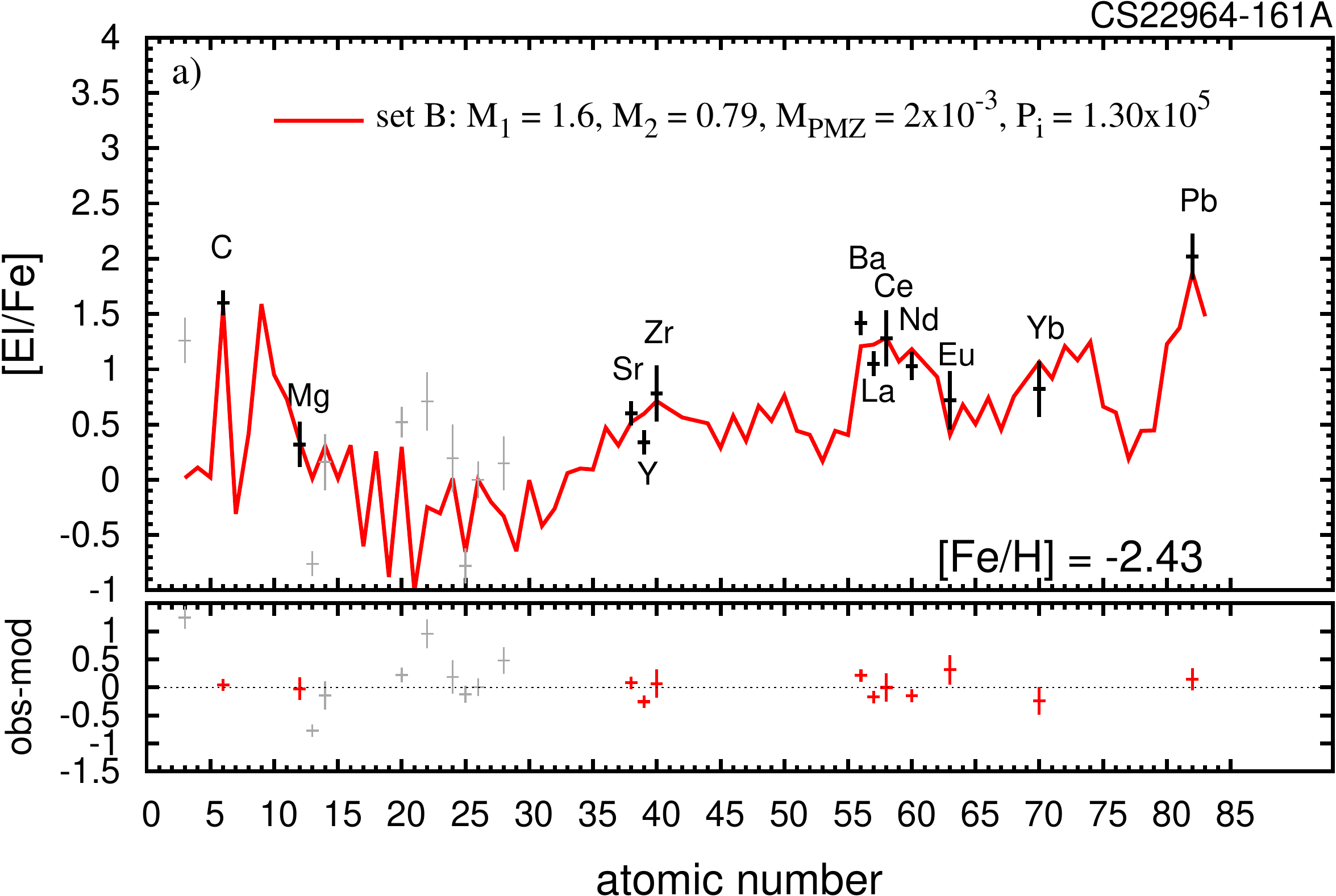}\hspace{5mm}
\includegraphics[angle=0, width=0.48\textwidth]{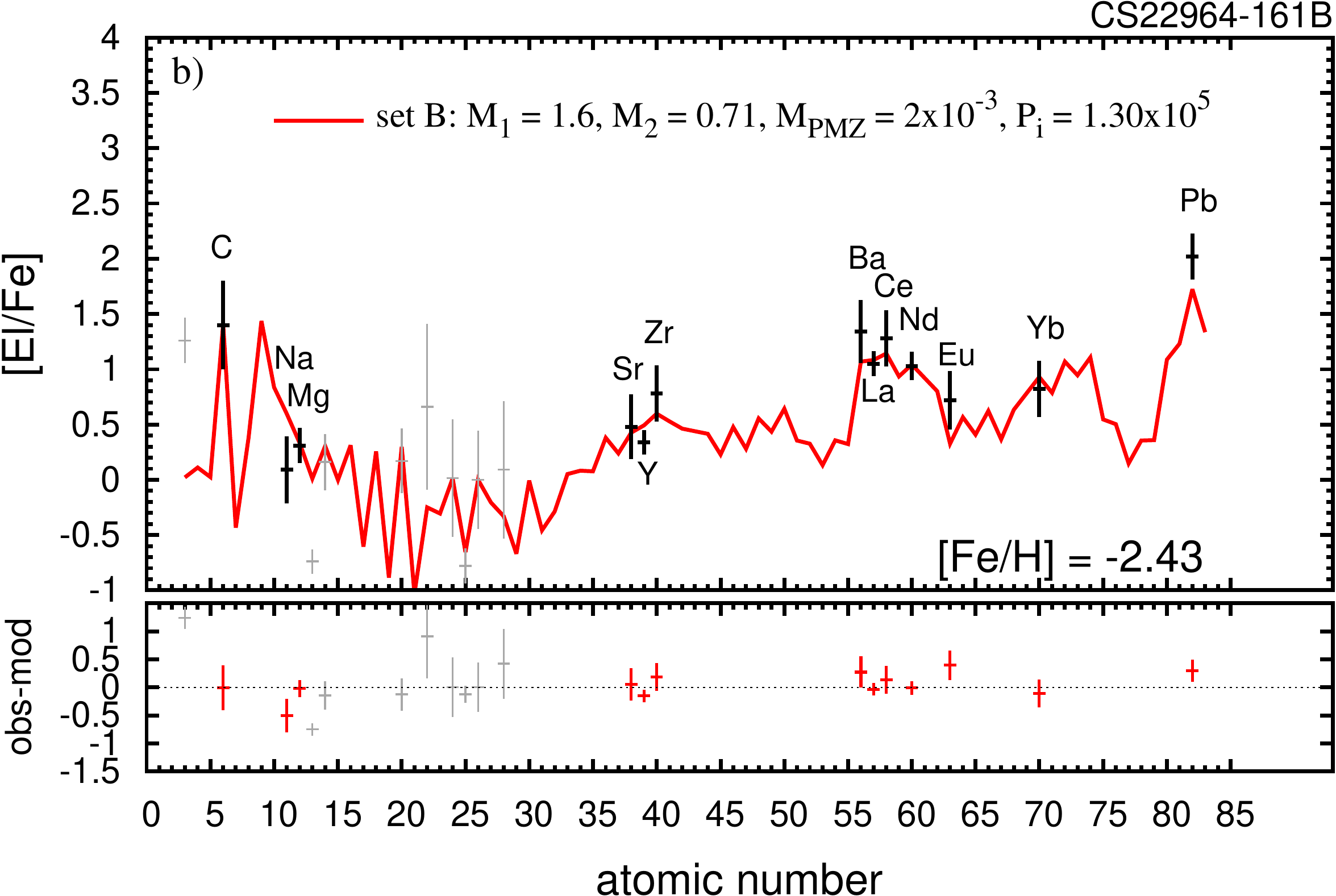}
\caption{Best-fitting models and observed abundances of stars CS22964--161A (left) and CS22964--161B (right). Symbols are the same as in Fig.~\ref{fig:CS22942-019}.}
\label{fig:CS22964-161}
\end{figure*}

The binary system CS22964--161A,B represents a special case, because both stars are enriched in carbon and $s$-elements and have relatively high surface gravities, consistent with stars that have recently passed the main-sequence turnoff (star A, Fig. \ref{fig:CS22964-161}a) or are still on the main sequence (star~B, Fig. \ref{fig:CS22964-161}b). The simplest interpretation is that the binary system seen today was the recipient of the material lost in the past from a third star in its AGB phase, in a hierarchical triple system \citep{Thompson2008}. In this hypothesis the observed orbital period ($\Porb=252$ days) does not correspond to the final period of our models, $\Pf$, which instead is the period of the conjectured, unseen third star. 

Our \texttt{binary\_c} code is suited to study the evolution of binary systems and not triple systems, hence to determine the best-fitting model to the observed abundances we use the following method. We find the models that best reproduce the abundances of stars A and B separately, with no restriction on the orbital period but with the constraint that the primary mass has to be the same for the two stars. We determine the initial masses $\Mprim$, $\Mpmz$, $M_{\mathrm{A,i}}$ and $M_{\mathrm{B,i}}$. We find that stars A and B accrete $0.04\Msun$ and $0.02\Msun$, respectively, and the total $\chimin$ is $\chisq_{\mathrm{min,\, A+B}}=31.4$ (with model set B). Subsequently we study the evolution of a binary system in which the initial primary and secondary masses are $\Mprim$ and $\Msec=M_{\mathrm{A,i}}+M_{\mathrm{B,i}}$, respectively. In this way we mimic the effect of a triple system where the primary star in a wide orbit ``sees'' the close inner binary as a single object with mass equal to the sum of the two components. We impose that the primary star transfers to the companion an amount of mass equal to $\Macc = 0.06\Msun$ and we determine the initial orbital period. In our best model star A is initially $10\%$ more massive than star B ($M_{\mathrm{A,i}}=0.79\Msun$ and $M_{\mathrm{B,i}}=0.71\Msun$). The final mass ratio $q=M_{\mathrm{B}}/M_{\mathrm{A}}$ is approximately $0.90$, similar to the value estimated from the observations \cite[$q=0.88$,][]{Thompson2008}. A choice of two stars with initially equal masses $M_A=M_B=0.76\,\Msun$ that accrete the same amount of material $\Delta\Macc\approx0.04\,\Msun$ leads to a model that is similar to our best model ($\chisq_{\mathrm{A+B}}=33$). 

Our best model reproduces the abundances of all the elements in both stars, except for sodium, which is overproduced in star B, and yttrium, which is slightly overestimated in both stars. We find equally good fits for both model sets, with the same parameters except for the initial and final periods of the third star.

%%%%%%%%%%%%%%%%%%%%%%%%%%%  SECTION 6  %%%%%%%%%%%%%%%%%%%%%%%%%%%
\section{Discussion}
\label{discussion}

\subsection{Constraints on mass transfer and binary evolution}
\label{disc_binev}

Our model set B (with enhanced Bondi-Hoyle-Lyttleton wind-accretion efficiency and efficient orbital angular momentum loss) provides the best fit to the observed abundances in ten stars and predicts results very similar to the best fit in four of the remaining five systems. This model set is not necessarily realistic because we assume an arbitrarily high efficiency of wind mass accretion. However, these results indicate that the accretion to the secondary star of large amounts of material is necessary in at least ten stars of our sample, otherwise we find a poor fit to the observations. 
Eight of these stars have low surface gravity, which implies that they evolved off the main sequence and have undergone the first dredge-up. Consequently, the accreted material is diluted throughout the envelope of the recipient star, regardless of our assumptions about the efficiency of thermohaline mixing. Large amounts of transferred mass are therefore required to reduce the dilution and reproduce the large enhancements of carbon and $s$-elements. 

Our model set A (with WRLOF wind-accretion efficiency and spherically symmetric wind) is disfavoured by the requirement of large amounts of mass accretion combined with the constraint on the orbital period. Because with a spherically symmetric wind the orbit typically expands in response to mass loss, the modelled binary stars need to be initially close. For orbital periods shorter than a few thousand days even a low-mass primary star transfers a small amount of material to the companion in our WRLOF model. Consequently, with model set A only stars BD$+04^{\circ}2466$ and CS22964--161A,B are well reproduced because the orbital periods are long and the observed carbon and $s$-elements are not strongly enhanced. In the future more realistic models of wind mass transfer are required to take into account more accurately the accretion process in close binary stars.

In six systems a common-envelope phase is necessary to shrink the orbit to the observed period (assuming model set B). The best models of five of these systems adopt the default common-envelope efficiency $\aCE=1$. Star HE0024--2523 (Fig. \ref{fig:HE0024-2523}) has an observed period of three days that can be reproduced only by assuming a very inefficient ejection process, $\aCE=0.03$.
An alternative interpretation is that HE0024--2523 was initially part of a hierarchical triple system in which an intermediate-mass primary star was in a wide orbit around two low-mass stars in a close binary. In this scenario the inner binary enters in a common envelope after being polluted by the primary star. Therefore, to reproduce the observed period an inefficient process of common-envelope ejection is not necessary. In an alternative triple scenario the unseen companion of HE0024--2523 is a low-mass main-sequence star and the inner binary was initially formed with the observed period of three days, in which case a common-envelope phase is not required. In summary, the $\aCE$ parameter is not well constrained and we do not find strong evidence to reject the default value $\aCE=1$.

\subsection{Comparison between modelled and observed abundances}

In some systems our model predictions do not reproduce the observed abundances even when we adopt a fine-tuned model of wind mass transfer. The abundance of nitrogen, for example, is determined in eight stars of our sample and it is well reproduced by our model only in two systems (HD201626 and HE0507--1430 in Figs. \ref{fig:HD201626} and \ref{fig:HE0507-1430}, respectively). However, as we noted in Sect. \ref{method} the exact amount of nitrogen produced by the CN cycle in AGB stars is uncertain. If some extra amount of carbon is converted to nitrogen in the model we reduce the discrepancies and in six out of eight stars the abundance of C+N is reproduced within the observational uncertainty.

The abundance of oxygen does not vary by more than approximately $1$ dex in the models of low-mass AGB stars and the models typically underestimate the observations. This discrepancy may indicate that some oxygen is mixed to the surface by the third dredge-up or that the initial abundance of oxygen adopted in our models is low compared to the average value observed in our sample of very metal-poor stars, as we will discuss in Paper II.

In many stars the observed abundances of sodium are low compared to the predictions (e.g. CS29497--030 and CS22964--161B). Sodium is produced in the intershell region by proton capture on neon seeds, therefore the sodium abundance grows very rapidly with increasing PMZ mass (see also the discussion about $^{23}\Na$ in K10 and \citealp{Lugaro2012}, and references therein). A better match would generally require a lower mass of the PMZ, but this is hard to reconcile with the large enhancements of $s$-elements that need a massive PMZ to be reproduced. The abundances of sodium in low-metallicity stars show a large dispersion which is difficult to reproduce in galactic chemical evolution models \cite[][]{Kobayashi2011}. However, our AGB models of mass above $\approx1.2\,\Msun$ produce large amounts of sodium and a factor of ten difference in its initial abundance accounts for no more than $0.1$ dex in the final [Na/Fe]. Therefore we consider it more likely that the discrepancy in sodium is related to the large uncertainties in the numerical treatment of the PMZ in the detailed models \cite[e.g.,][]{GorielyMowlavi2000, Lugaro2004}.

In CS22964--161A,B the observed abundances are generally well reproduced, but yttrium is several tenths of a dex lower than strontium and zirconium. Such a distribution of  strontium, yttrium and zirconium is observed in other CEMP-$s$ stars, as we will discuss in Paper II, but it is not replicable by our models because these elements are always produced in comparable amounts. The observed abundance of zirconium is low compared to the models of the stars HD198269, HD$201626$ and HD224959 (Figs. \ref{fig:HD198269}--\ref{fig:HD224959}). This discrepancy may be related with a systematic effect in the observations by \cite{VanEck2003}, because previous measurements from spectra at lower resolution by \cite{Vanture1992a,Vanture1992b} are consistent with the models.

\subsection{Abundances in CEMP-s/r stars}
The CEMP-$s/r$ stars in our sample, CS22948--027 (Fig. \ref{fig:CS22948-027}), CS29497--030 (Fig. \ref{fig:CS29497-030}), CS29497--034 (Fig.  \ref{fig:CS29497-034}), HD224959 (Fig. \ref{fig:HD224959}) and LP625--44 (Fig. \ref{fig:LP625-44}), show highly enriched abundances of elements that are mostly produced in the $r$-process, such as europium. These elements are typically produced in small amounts in our AGB nucleosynthesis model. Consequently, to reproduce the large enhancements of the $r$-elements our models often overestimate the abundances of other elements, such as carbon, sodium and light-$s$ elements.
This issue illustrates the general problem of reproducing element-to-element ratios. More specifically, in some systems a high value of $\Mpmz$ is necessary to match one element whereas a lower $\Mpmz$ is sufficient to reproduce another element. In LP625--44, for example, all the elements in the heavy-$s$ peak are underestimated and would need a higher $\Mpmz$ (or a higher $M_1$) to be reproduced. On the contrary light elements, light-$s$ elements and lead are overestimated and would be better reproduced with a lower $\Mpmz$ or $M_1$. 

The difficulty in reproducing the abundance ratios suggests that our simple parameterisation of the PMZ possibly ignores some important aspects of the physics involved in the problem. For example, the mass of our PMZ stays constant during the evolution of the AGB. In the neutron-capture process the $s$-element peaks are filled progressively in time: in the first few thermal pulses the star produces mostly light-$s$ elements while in later thermal pulses the distribution is weighed more towards neutron-rich nuclei. A time-dependent $\Mpmz$ could modify the final light-$s$ to heavy-$s$ ratios. For example, a PMZ that increases in time may be able to produce larger $[\hs/\ls]$ and $[\Pb/\hs]$ ratios. Values of [hs/ls] above 1 dex are observed in most CEMP-$s/r$ stars and are currently impossible to reproduce in our models, as we will discuss in Paper II. On the other hand, a small PMZ in the last few pulses would probably enhance the light-$s$ peak compared to the heavy-$s$ peak and lead. 

The origin of the $r$-element enrichment in CEMP-$s$ stars is an open issue. Several explanations have been proposed and some of these suggest that the $r$- and $s$-enrichments in CEMP-$s/r$ stars are independent \citep[see][]{Jonsell2006}. However, this interpretation does not explain the correlation observed in CEMP-$s$ stars between the abundances of barium and europium. 

\subsection{The effect of non-convective mixing in dwarf CEMP stars}
Five stars in our sample have high surface gravities and therefore have not yet experienced the first dredge-up. The mixing mechanisms that can modify the surface abundances in these stars are not well understood and many counteracting effects potentially play a role: thermohaline mixing, gravitational settling, radiative levitation and rotation. In our models we simulate two opposite situations: our default option is that the accreted material is diluted throughout the entire star. Alternatively, the mixing is inefficient and the material remains on the surface until mixed in by convection. The choice of inefficient mixing improves the fit of two stars, CS22956--028 and CS29497--030. However, the fit of star CS22956--028 (Fig. \ref{fig:CS22956-028}) is based on only five elements one of which, barium, is overestimated by 2 dex. An alternative interpretation of the observed abundances is that this star may have been polluted by a rotating massive star, which may produce up to $[\Sr/\Ba]\approx2$ and $[\Pb/\Sr]\lesssim-1$ as suggested by the models of $25\Msun$ stars discussed by \cite{Frischknecht2012}. The abundances of other elements, particularly nitrogen and lead, are necessary to test this hypothesis. On the other hand, CS29497--030 is a CEMP-$s/r$ star and the $\chisq$ of the fit is dominated by the large abundance of seven elements mainly produced by the $r$-process. A fit that includes efficient thermohaline mixing reproduces the $s$-process elements much better. In summary, our results do not provide strong constraints on the efficiency of non-convective mixing mechanisms in metal-poor dwarf stars.

\subsection{The effect of fixed model metallicity}

The fact that we use a model tailored for $[\Fe/\Hy]=-2.24$ to investigate stars of different iron abundances is an issue for many stars in our sample. By adopting a fixed metallicity for all our systems we implicitly ignore the effects due to the different $[\Fe/\Hy]$. This approximation likely introduces increasingly bigger errors the larger the difference in $[\Fe/\Hy]$.
Qualitatively, we expect the choice of a lower metallicity to result in two effects: an increase in the abundances of carbon and all the elements produced by AGB nucleosynthesis, because the iron abundance is smaller, and relatively larger abundances of neutron-rich isotopes, because of the higher ratio of neutrons to iron seeds. If we vary the metallicity in our models, all other options being equal, we reproduce the first of the two effects but not the second, because the neutron-to-seed ratio, and consequently the amount of $s$-process elements that are produced in our modelled intershell region, is determined by the mass of the star and by the mass of the PMZ at $[\Fe/\Hy]=-2.24$. Hence, we need to assume that the abundances of $s$-isotopes in the intershell region do not change for metallicity variations.
It is arguable which of the two options is the best: to adopt the same metallicity for all observed systems or to vary it in proportion to the observed $[\Fe/\Hy]$.
In this study we preferred the first option because we are more confident of the results of our code at the fixed metallicity $Z=10^{-4}$, but it is very important to keep in mind the limitations that derive from this choice. 

The most extreme example is CEMP-$s/r$ star CS29497--034 that exhibits iron abundance $[\Fe/\Hy]=-2.96$ which corresponds to a metallicity five times lower than our default assumption. However, if we adopt $Z=10^{-4}$ we find a model that reproduces the abundances of all elements except europium (solid line in Fig. \ref{fig:CS29497-034}), while if we adopt $Z=2\times10^{-5}$ our best model overestimates the abundances of carbon, sodium and magnesium (dashed line in Fig. \ref{fig:CS29497-034}) and does not improve the fit to the abundance of europium. This result suggests that the intershell composition of AGB stars at $[\Fe/\Hy] \approx -3$ may be very different. 

Another example is HE0024--2523, that exhibits a distribution of $s$-elements weighted towards neutron-rich elements and none of our model reconciles the large abundance of lead, $[\Pb/\Fe]=3.2$, with the low enhancement of strontium, $[\Sr/\Fe]=0.5$ (Fig. \ref{fig:HE0024-2523}). This discrepancy can be qualitatively explained by the difference in the observed and modelled metallicity ($Z\approx4\times10^{-5}$ and $Z=10^{-4}$, respectively). However, star CS22942--019 has almost the same iron abundance ($[\Fe/\Hy]=-2.69$) but we find a good match between our best model and the observations.

%%%%%%%%%%%%%%%%%%%%%%%%%%%  SECTION 7  %%%%%%%%%%%%%%%%%%%%%%%%%%%
\section{Conclusions}
\label{conclusions}

This study shows that the requirement to reproduce at the same time the chemical abundances and orbital periods observed in our sample of $15$ CEMP-$s$ binary stars put strong constraints on the adopted binary evolution model. It is generally necessary that the modelled binary systems lose efficiently angular momentum and transfer mass with high accretion efficiency. In particular, binary systems with orbital periods below a few thousand days need to transfer mass more efficiently than normally assumed in our models. In a forthcoming paper we will analyse a larger sample of CEMP-$s$ stars without measured orbital periods. The comparison between the modelled periods determined in the two studies will have implications for our model of the wind mass-transfer process.

When the condition of strong mass transfer is fulfilled the discrepancies between synthetic and observed abundances arise from the model of AGB nucleosynthesis. 
In about half of the systems the observed element-to-element ratios are not reproduced. In particular, to match the large enhancements of the heavy-$s$ elements in CEMP-$s/r$ stars our models produce an excess of carbon, sodium, magnesium and light-$s$ elements. This discrepancy and the fact that the abundances of the $r$-process elements are mostly underestimated suggest that in our model higher densities of free neutrons should be produced in some circumstances, because a larger neutron-to-iron ratio would favour neutron-rich elements.

%%%%%%%%%%%%%%%%%%%%%%%%%%%
\begin{acknowledgements}
We are grateful to dr. T. Suda for his kind support and help to access the SAGA database. We thank the referee for his comments that have helped to improve the clarity and conciseness of this paper. CA is grateful for financial support from the Netherlands Organisation for Scientific Research (NWO) under grant 614.000.901. 
\end{acknowledgements}

%%%%%%%%%%%%%%%%%%%%%%%%%%%  BIBLIOGRAPHY  %%%%%%%%%%%%%%%%%%%%%%%%%%

%%%%%%%%%%%%%%%%%%%%%%%%%%%%%%%%%%%%%%%%%%%%%%%%%%%%%

\Online
\begin{appendix}
%%%%%%%%%%%%%%%%%%%%%%%%%%%  APPENDIX A  %%%%%%%%%%%%%%%%%%%%%%%%%%
\section{Best-fitting models to the observed CEMP-$s$ stars in our sample}
\label{app:A}
In this section we show the best-fitting models to the observed abundances of all CEMP-$s$ stars in our sample.

%%%%%%%%%%%%%%%%%%%%%%%%%%%%%%%
\begin{figure}[!h]
\includegraphics[angle=0, width=0.488\textwidth]{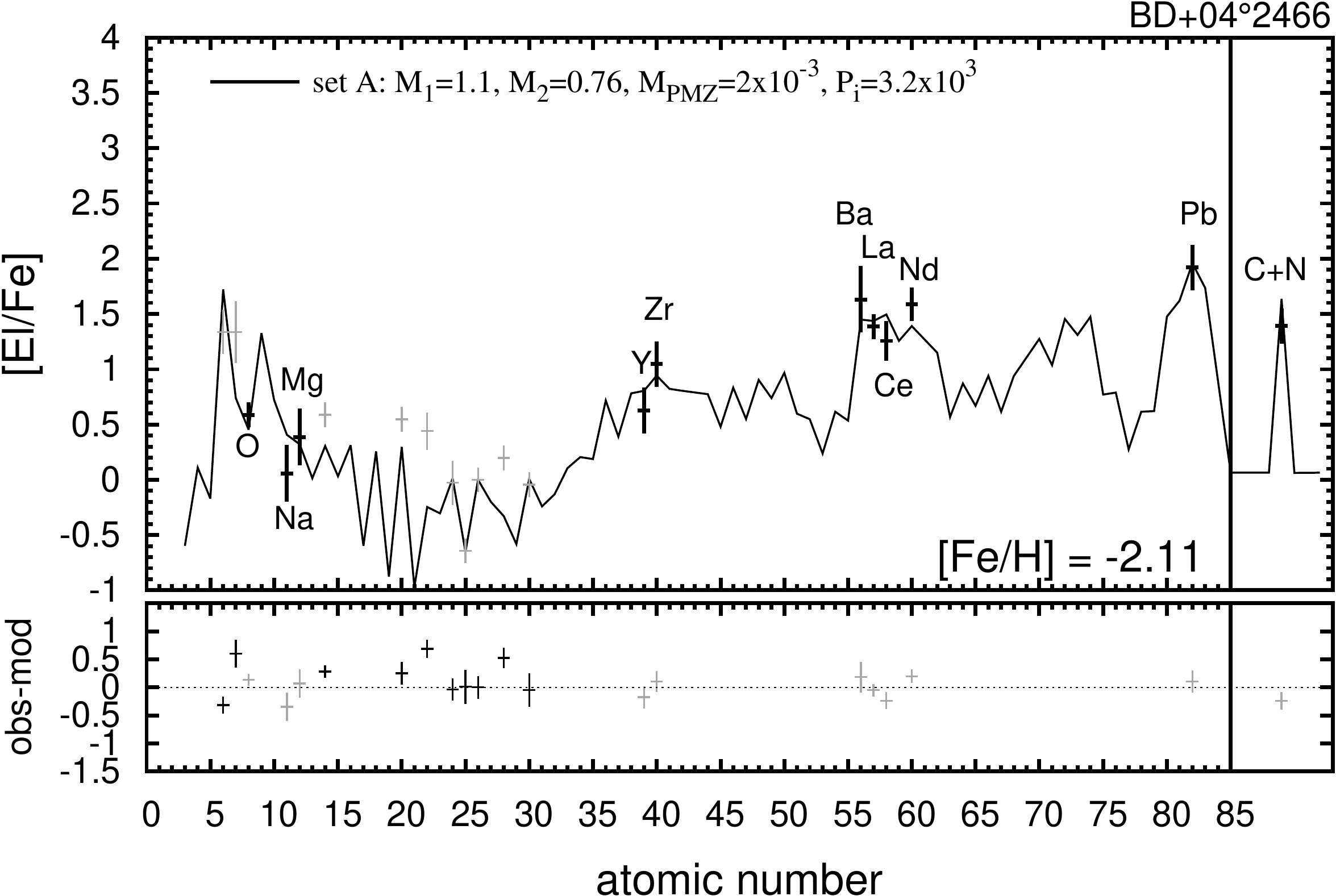}
\caption{\footnotesize{{\it Points with error bars}: as Fig. \ref{fig:CS22942-019} for BD$+04^{\circ}2466$. {\it Solid line}: best-fitting model computed with model set A (Table \ref{tab:bestfit_WRLOFq}). The combined abundance of carbon and nitrogen is shown in the right panel.}}
\label{fig:BD+04o2466}
\end{figure}
%

%%%%%%%%%%%%%%%%%%%%%%%%%%%%%%%%%%%%
\begin{figure}[!h]
\vspace{15mm}
\includegraphics[angle=0, width=0.488\textwidth]{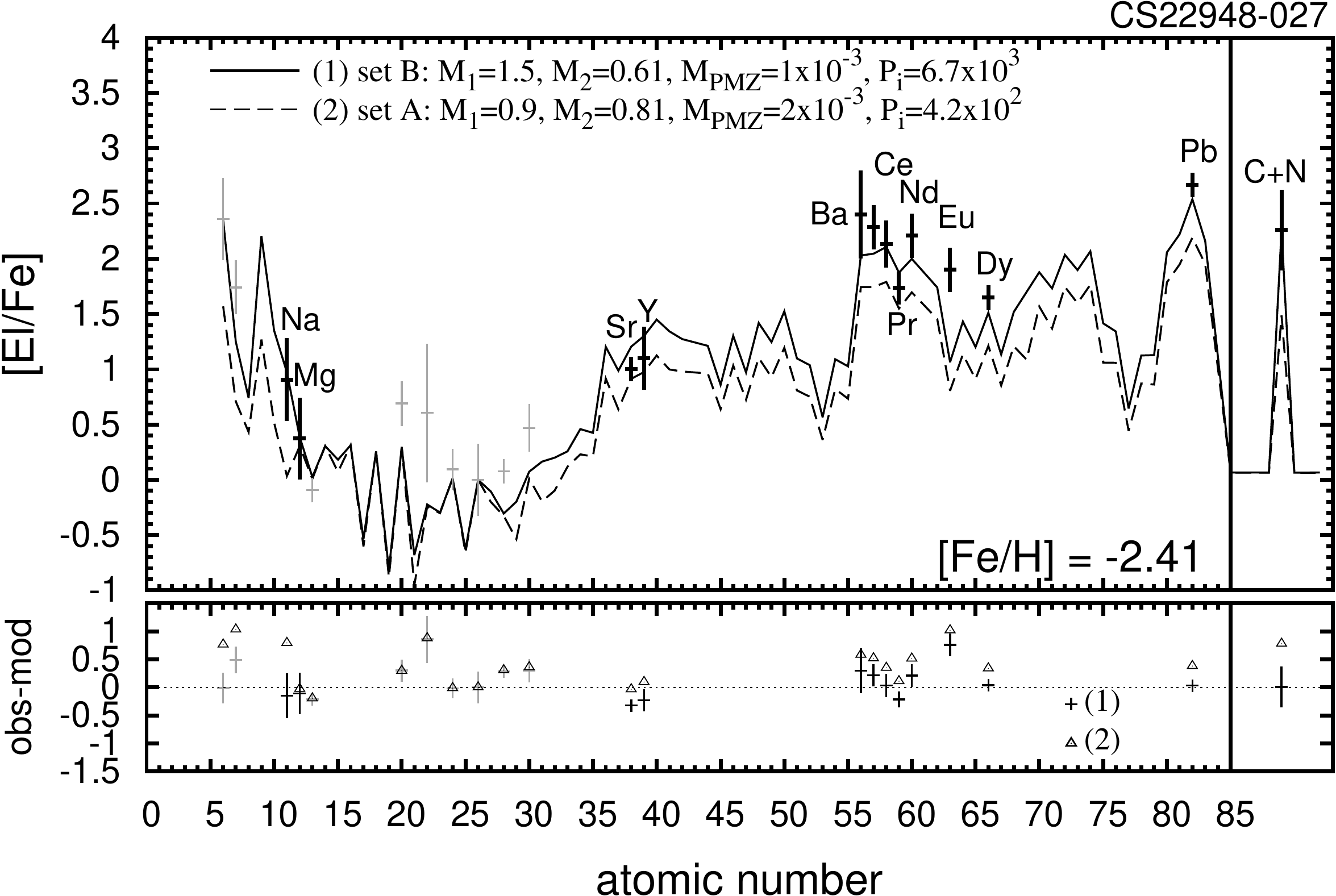}
\caption{\footnotesize{Star CS$22948-027$. {\it Solid line and points}: as Fig. \ref{fig:BD+04o2466}. In the best model (with model set B) the secondary star accretes $\Delta\Macc=0.27\,\Msun$. With model set A ({\it dashed line}) the secondary star accretes $\Delta\Macc=0.12\,\Msun$ and therefore the material is more strongly diluted.}}
\label{fig:CS22948-027}
\end{figure}
%

%%%%%%%%%%%%%%%%%%%%%%%%%%%%%%%
\begin{figure}[!h]
\includegraphics[angle=0, width=0.48\textwidth]{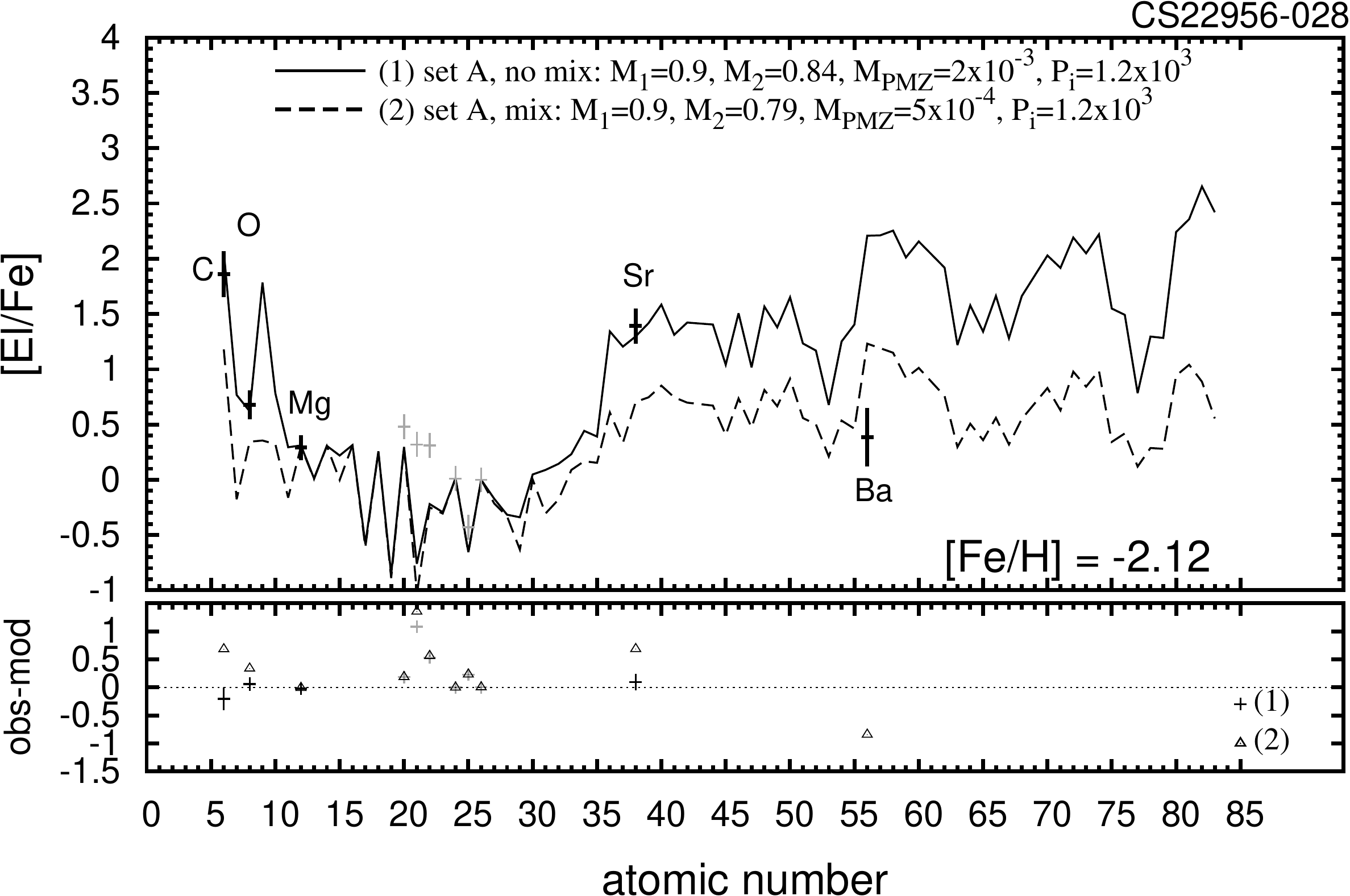}
\caption{\footnotesize{{\it Solid line and points}: as Fig. \ref{fig:BD+04o2466} for CS22956--028. The best-fitting model assumes that non-convective mixing processes, such as thermohaline mixing, are inefficient and therefore the accreted material remains on the surface of the star. {\it Dashed line}: alternative model with efficient thermohaline mixing ($\chisq=51.8$).}}
\label{fig:CS22956-028}
\end{figure}
%

%%%%%%%%%%%%%%%%%%%%%%%%%%%%%%%%%%%%
\begin{figure}[!h]
\includegraphics[angle=0, width=0.488\textwidth]{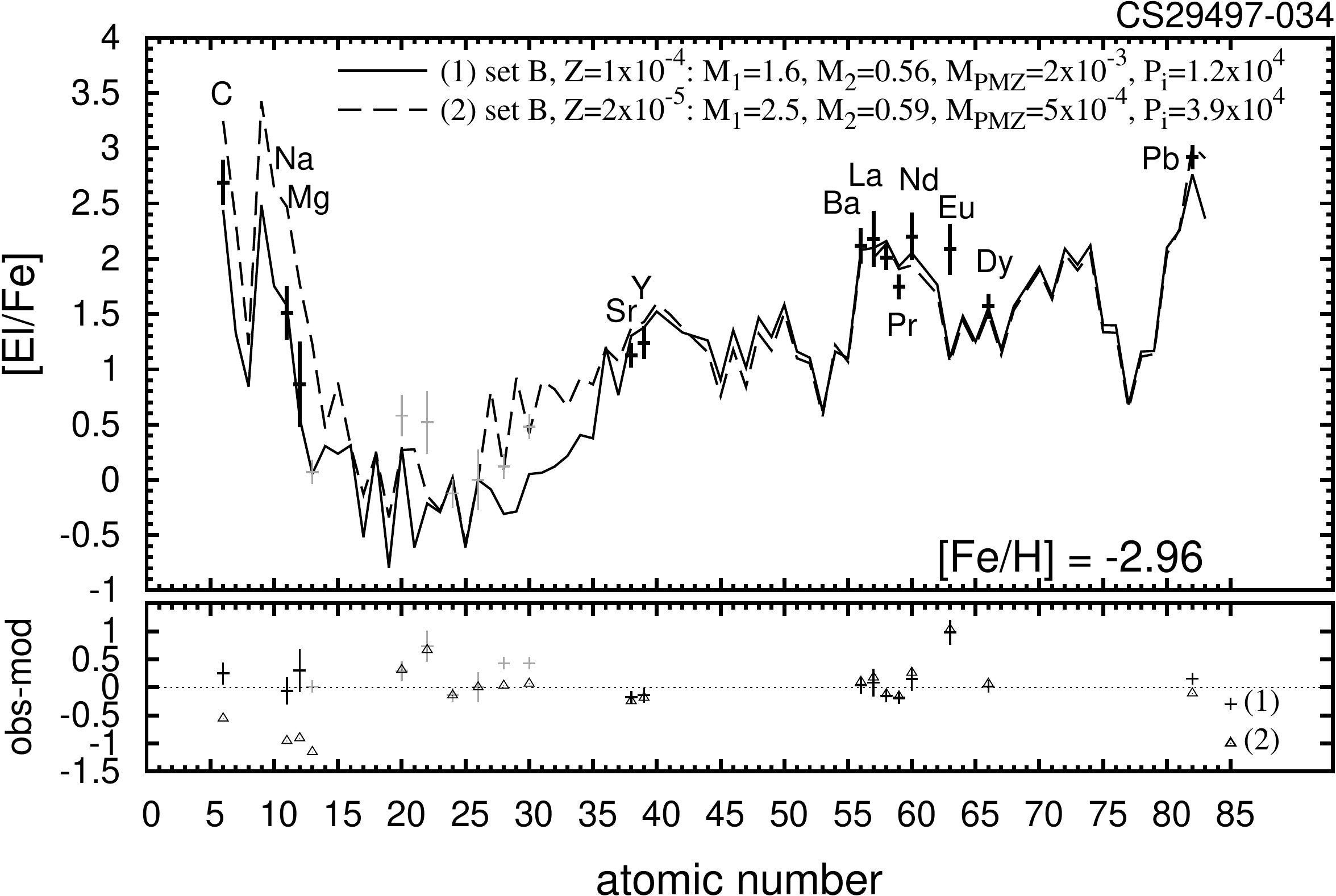}
\caption{\footnotesize{{\it Points with error bars}: as in Fig.~\ref{fig:BD+04o2466} for CS29497--034. {\it Solid line}: best-fitting model computed with set B adopting the default metallicity $Z=10^{-4}$. $\chimin$ is dominated by the discrepancy in the abundance of europium (about 1 dex). If we exclude europium from the fit we find $\chimin=15.2$ and $\chimin/\nu=1.7$. {\it Dashed line}: alternative model computed with set B and reduced metallicity, $Z=2\times10^{-5}$, that corresponds to the observed iron abundance $[\Fe/\Hy]=-2.96$.}}
\label{fig:CS29497-034}
\end{figure}
%

%%%%%%%%%%%%%%%%%%%%%%%%%%%%%%%%%%%%
\begin{figure}[!b]
\includegraphics[angle=0, width=0.48\textwidth]{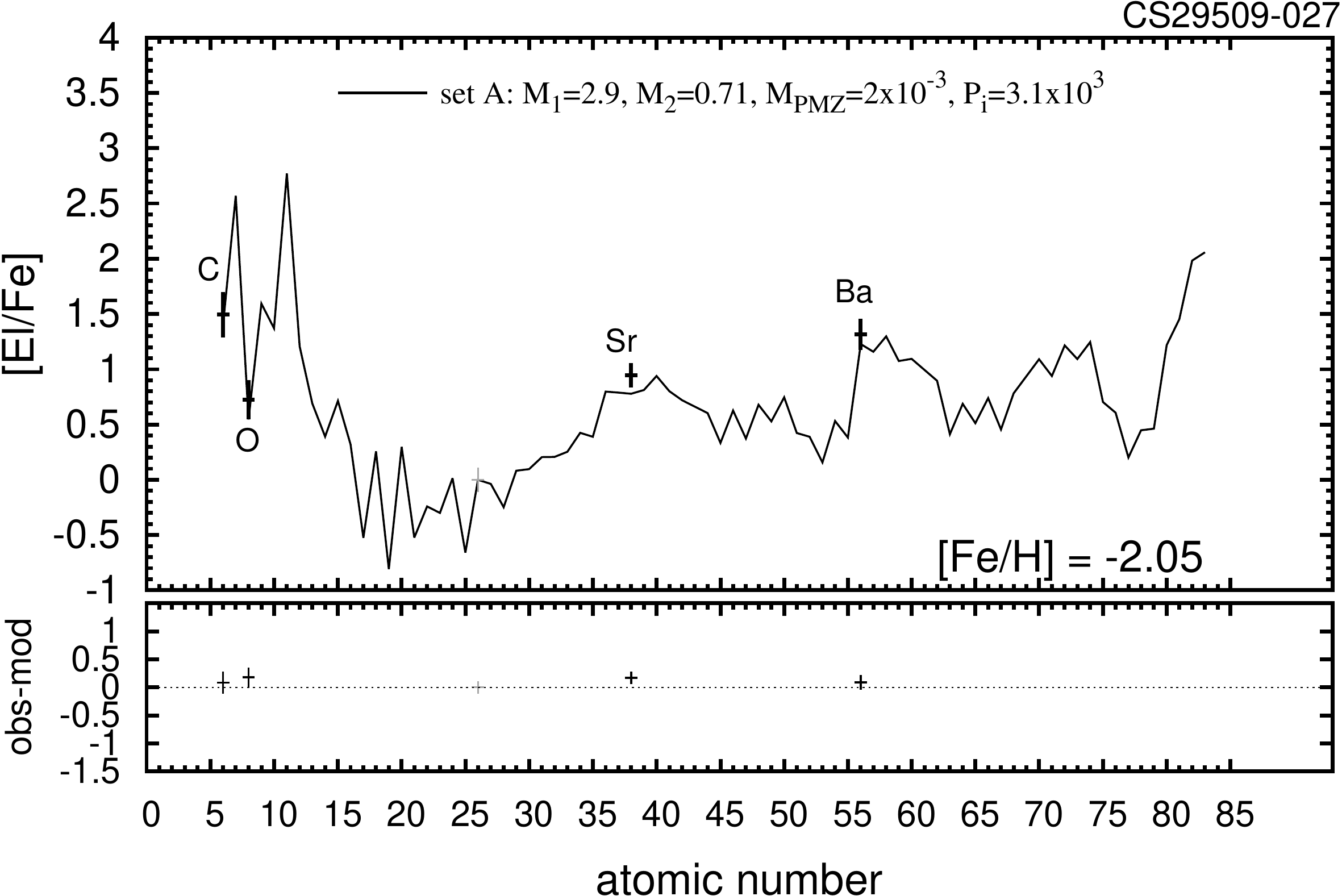}
\caption{\footnotesize{As Fig. \ref{fig:BD+04o2466} for CS29509--027. The abundance of other elemens, such as N, Na, Mg and Pb, is necessary to better constrain the initial primary mass.}}
\label{fig:CS29509-027}
\end{figure}
%

%%%%%%%%%%%%%%%%%%%%%%%%%%%%%%%%%%%%
\begin{figure}[!b]
\includegraphics[angle=0, width=0.48\textwidth]{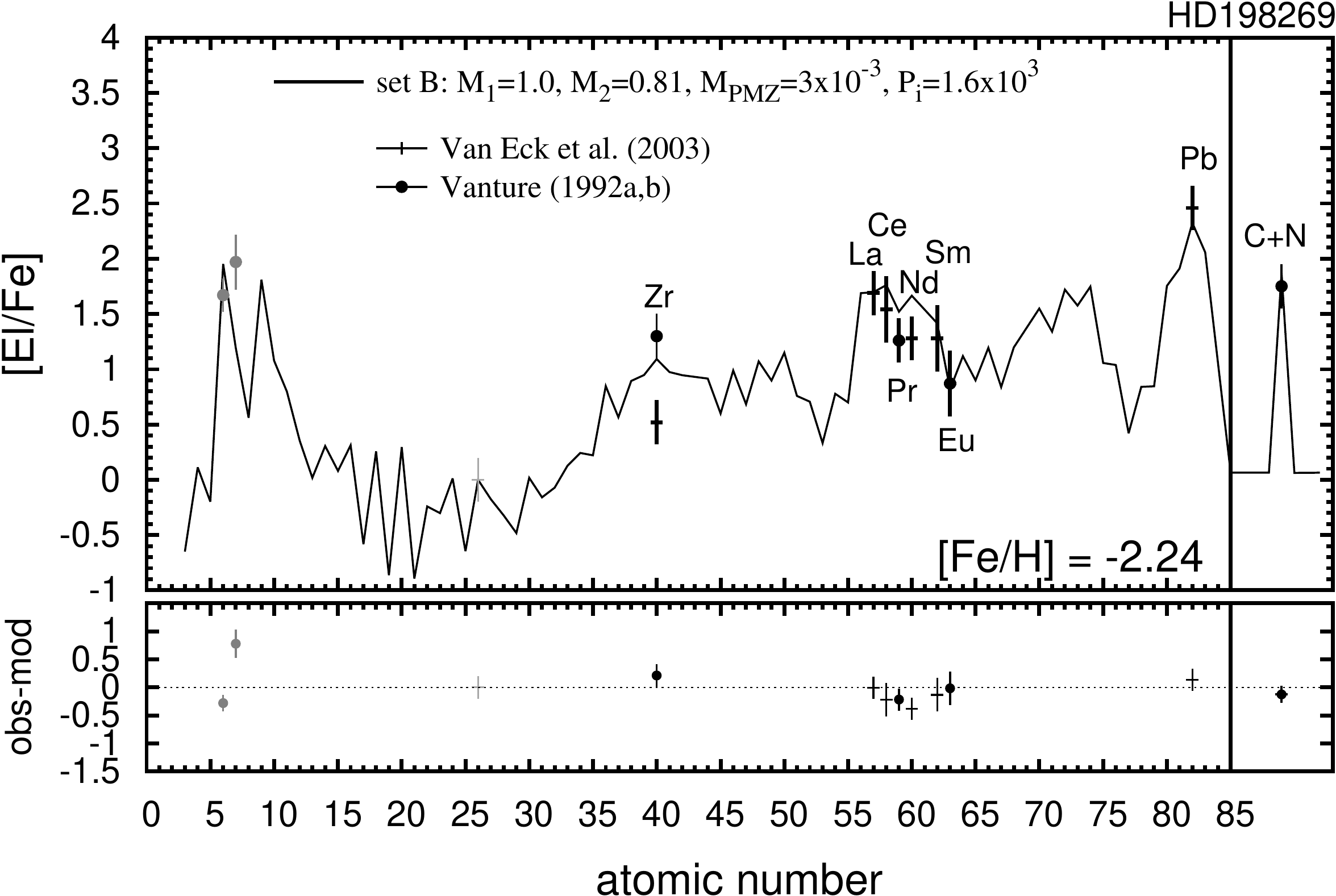}
\caption{\footnotesize{As Fig. \ref{fig:BD+04o2466} for HD198269. The best-fitting model ({\it solid line}) is computed adopting model set B. The abundances determined by \cite{VanEck2003} and \cite{Vanture1992b} are represented as plus signs and filled circles, respectively.}}
\label{fig:HD198269}
\end{figure}
%

%%%%%%%%%%%%%%%%%%%%%%%%%%%%%%%%%%%%
\begin{figure}[!b]
\includegraphics[angle=0, width=0.48\textwidth]{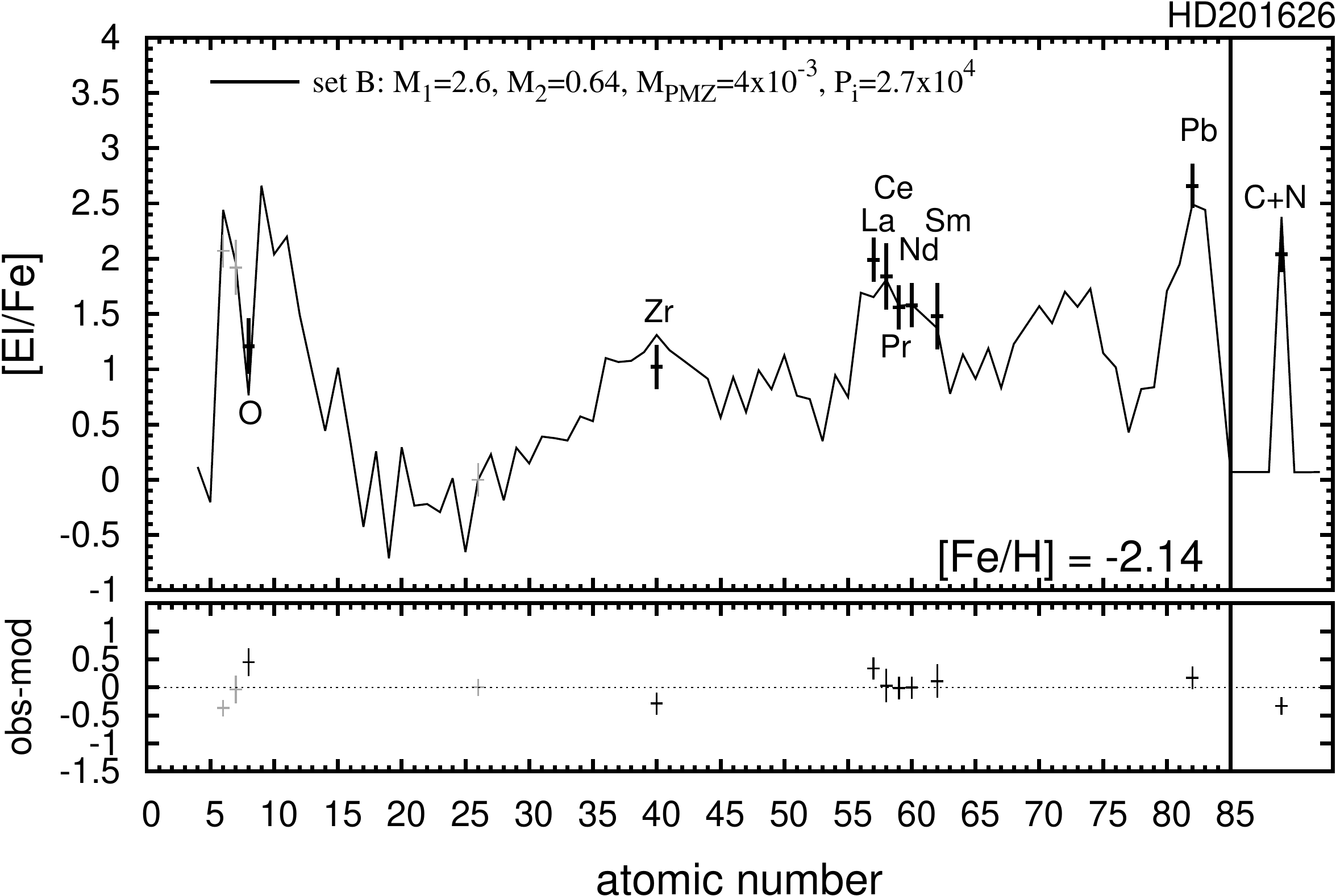}
\caption{\footnotesize{As Fig. \ref{fig:BD+04o2466} for HD201626. The best-fitting model to the observed abundances is found with model set B.}}
\label{fig:HD201626}
\end{figure}
%

%%%%%%%%%%%%%%%%%%%%%%%%%%%%%%%%%%%%
\begin{figure}[!b]
\includegraphics[angle=0, width=0.48\textwidth]{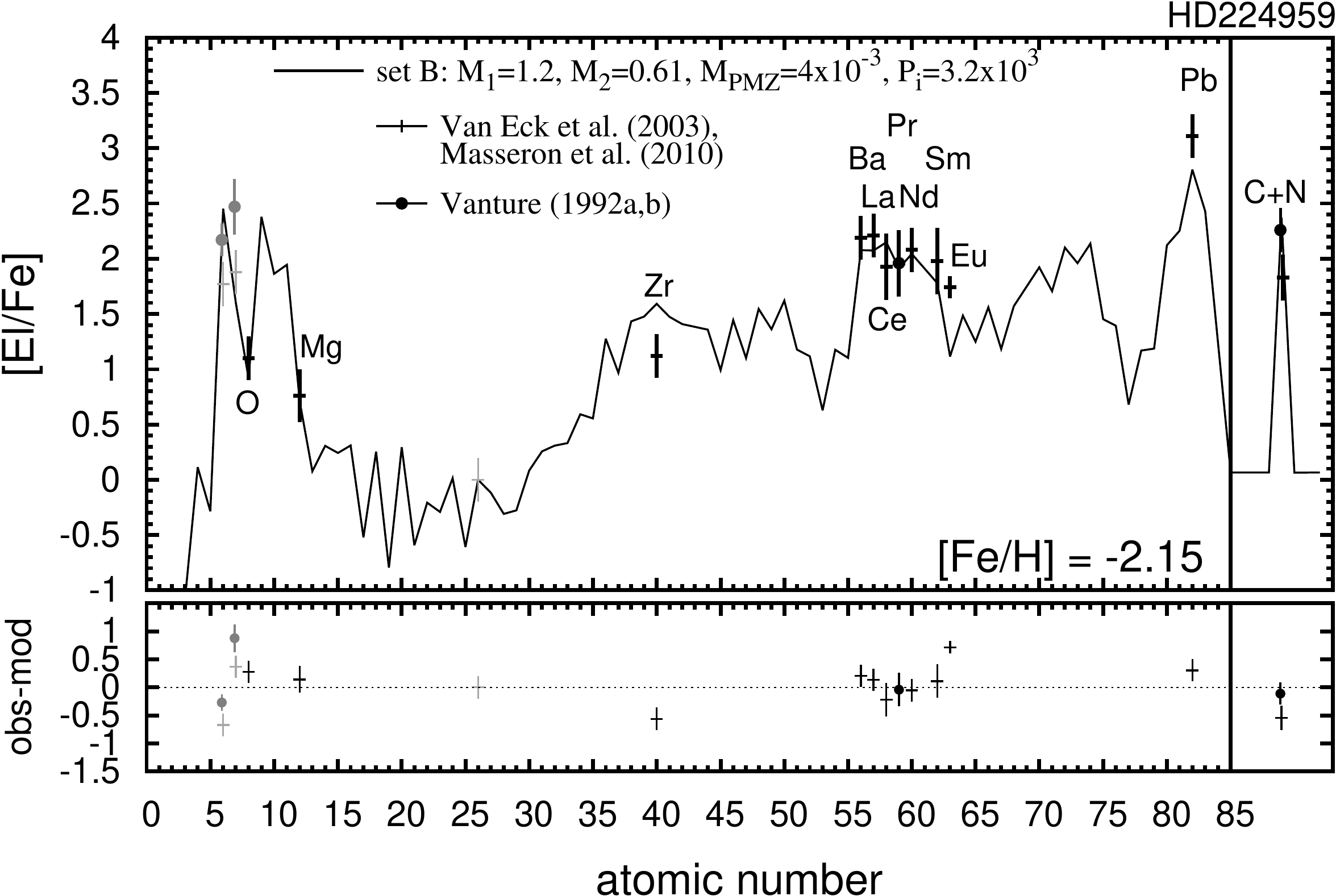}
\caption{\footnotesize{As Fig. \ref{fig:BD+04o2466} for CEMP-$s/r$ star HD224959. The abundances determined by \cite{VanEck2003} and \cite{Vanture1992b} are represented as plus signs and filled circles, respectively.}}
\label{fig:HD224959}
\end{figure}
%

%%%%%%%%%%%%%%%%%%%%%%%%%%%%%%%%%%%%
\begin{figure}[!b]
\includegraphics[angle=0, width=0.488\textwidth]{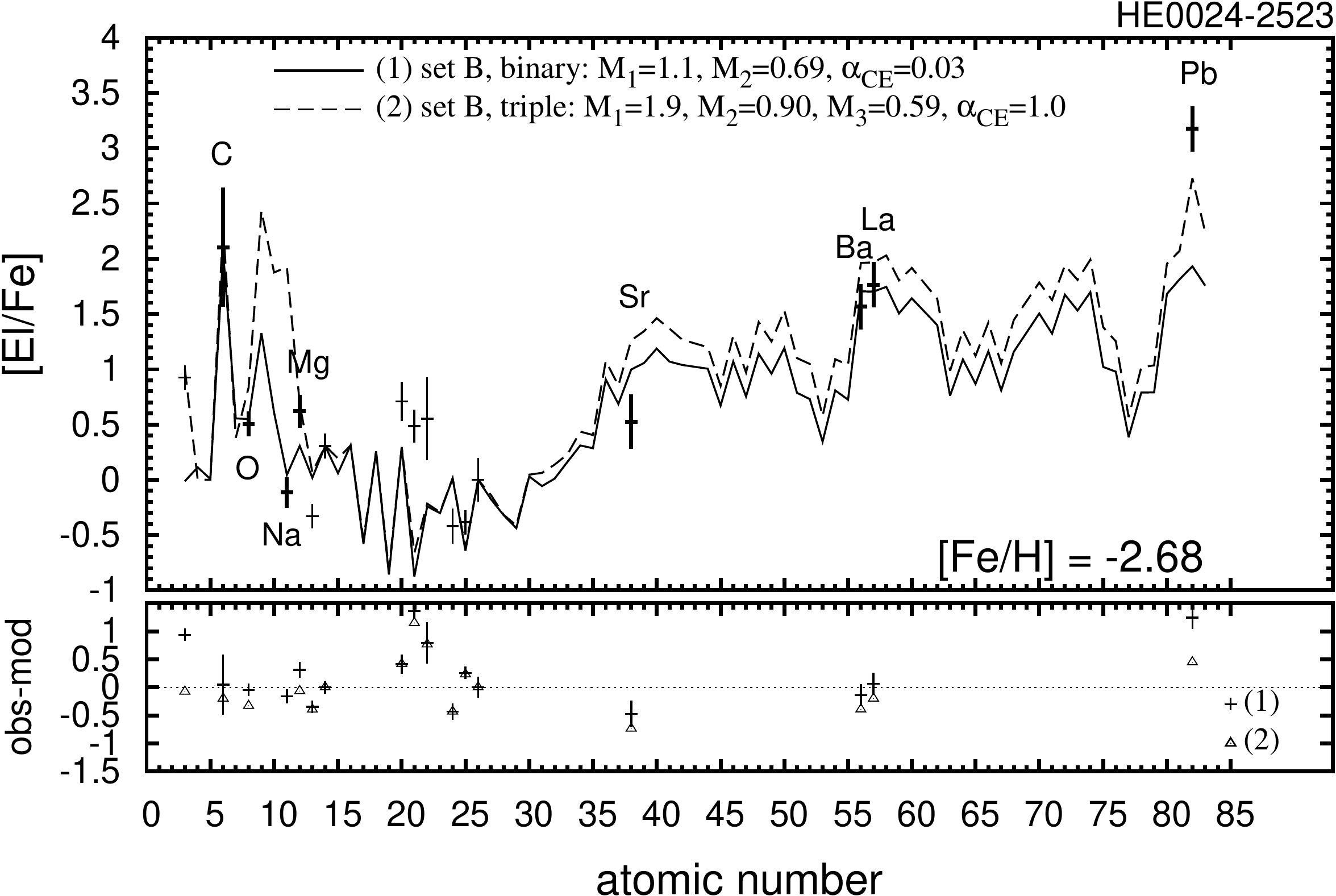}
\caption{\footnotesize{{\it Points with error bars:} as Fig. \ref{fig:BD+04o2466} for HE0024--2523. {\it Solid line:} Best-fitting model found with model set B and assuming a low efficiency for common-envelope ejection, $\aCE=0.03$, which is required to reproduce the observed period of $\Porb=3.14$ days in a binary scenario. {\it Dahsed line:} alternative scenario in which the HE0024--2523 was initially part of a hierarchical triple system. An intermediate-mass primary star was in a wide orbit around a close binary. During its AGB phase the primary star pollutes the inner binary. Subsequently, the secondary star overfills its Roche lobe, the system enters a common envelope the ejection of which (modelled with $\aCE=1.0$) requires the orbit to shrink to the observed period.}}
\label{fig:HE0024-2523}
\end{figure}
%

%%%%%%%%%%%%%%%%%%%%%%%%%%%%%%%
\begin{figure}[!t]
\includegraphics[angle=0, width=0.48\textwidth]{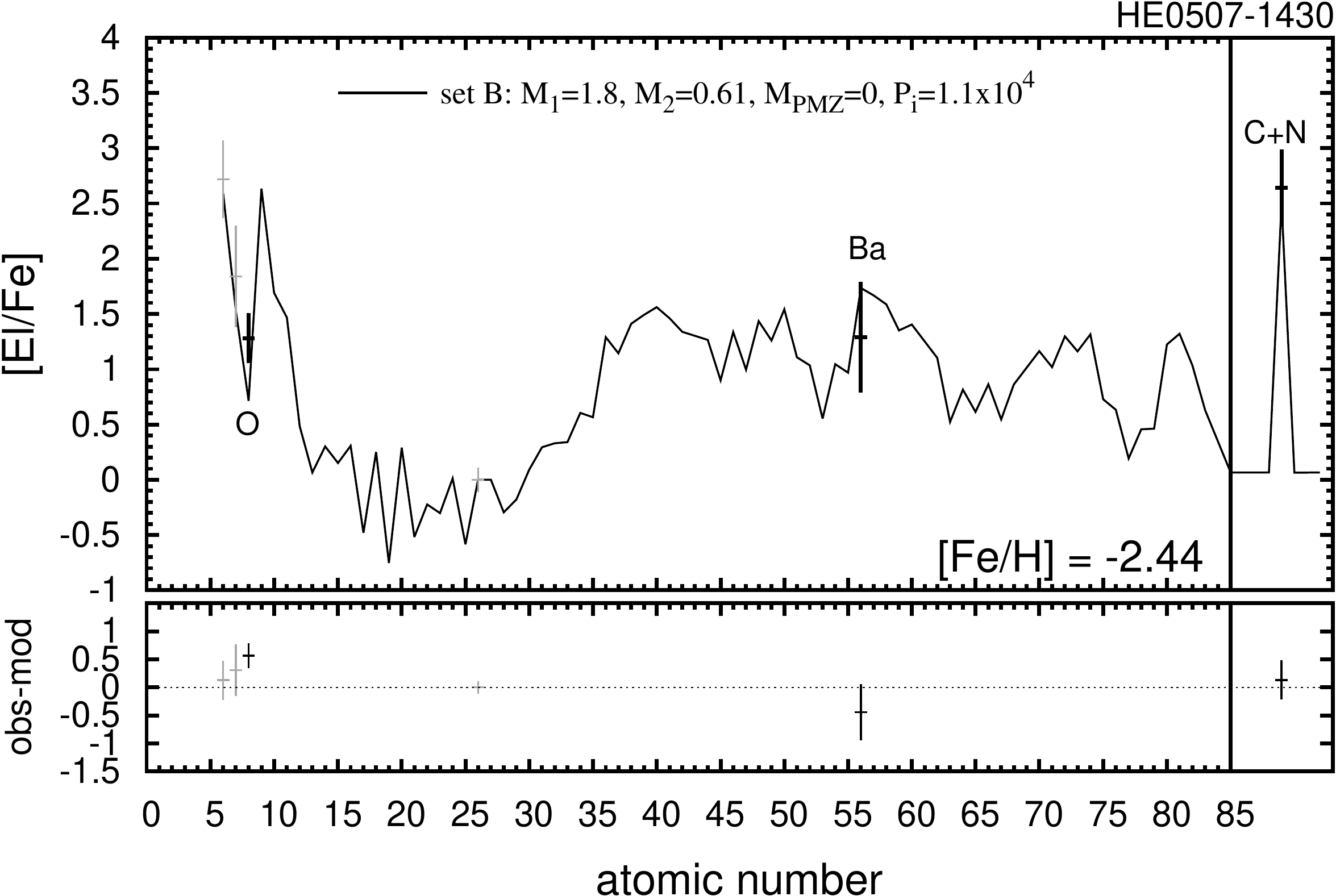}
\caption{\footnotesize{As Fig. \ref{fig:BD+04o2466} for star HE$0507-1430$. The best-fit model is found with model set B.}}
\label{fig:HE0507-1430}
\end{figure}
%

%%%%%%%%%%%%%%%%%%%%%%%%%%%%%%%
\begin{figure}[!t]
\includegraphics[angle=0, width=0.488\textwidth]{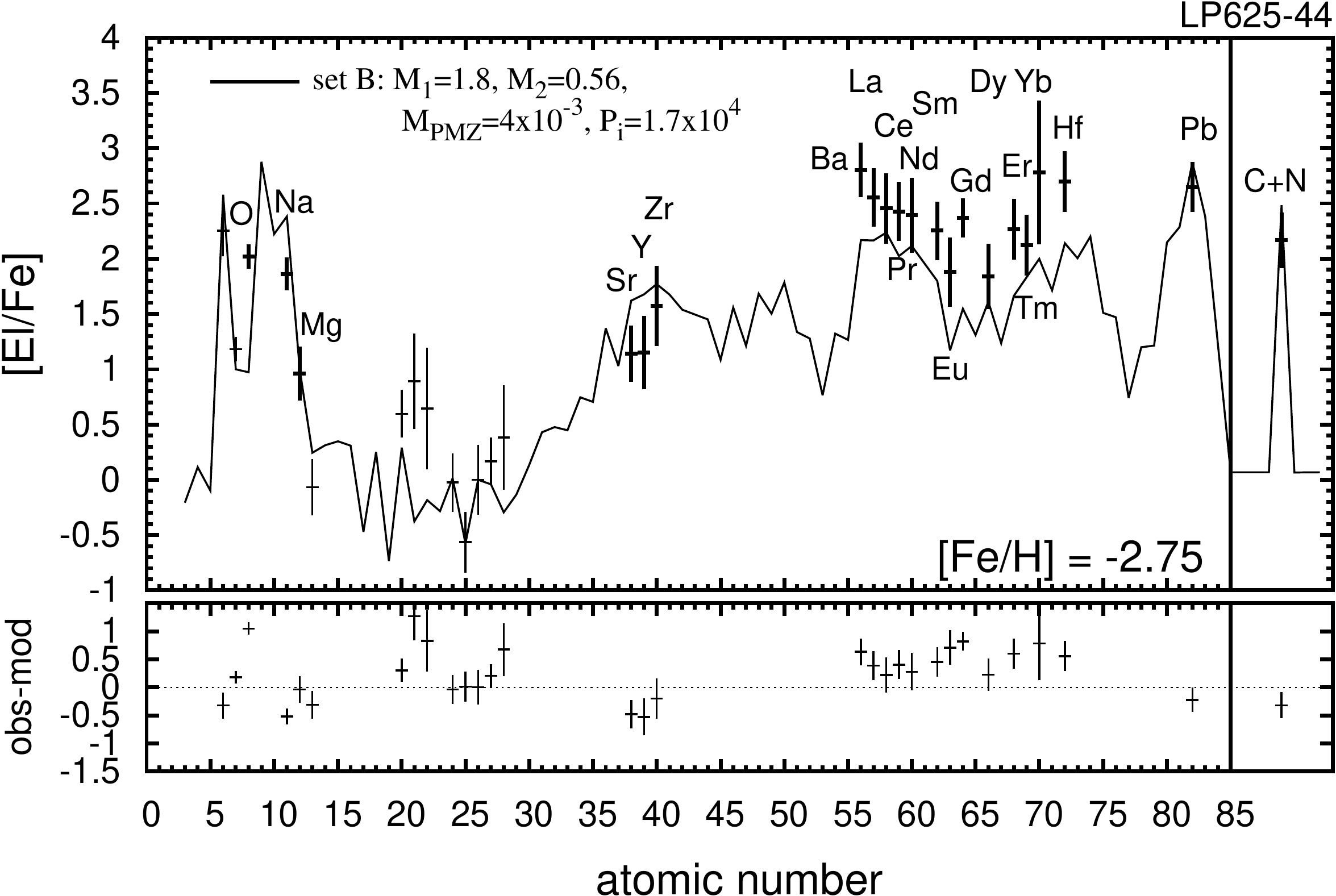}
\caption{\footnotesize{As in Fig. \ref{fig:BD+04o2466} for star LP$625-44$. The best-fit model is found with model set B.}}
\label{fig:LP625-44}
\end{figure}

\clearpage

%%%%%%%%%%%%%%%%%%%%%%%%%%%  APPENDIX B  %%%%%%%%%%%%%%%%%%%%%%%%%%
\section{Two-dimensional confidence intervals}
\label{app:B}
\begin{figure}[!h]
\centering
\includegraphics[angle=0, width=0.48\textwidth]{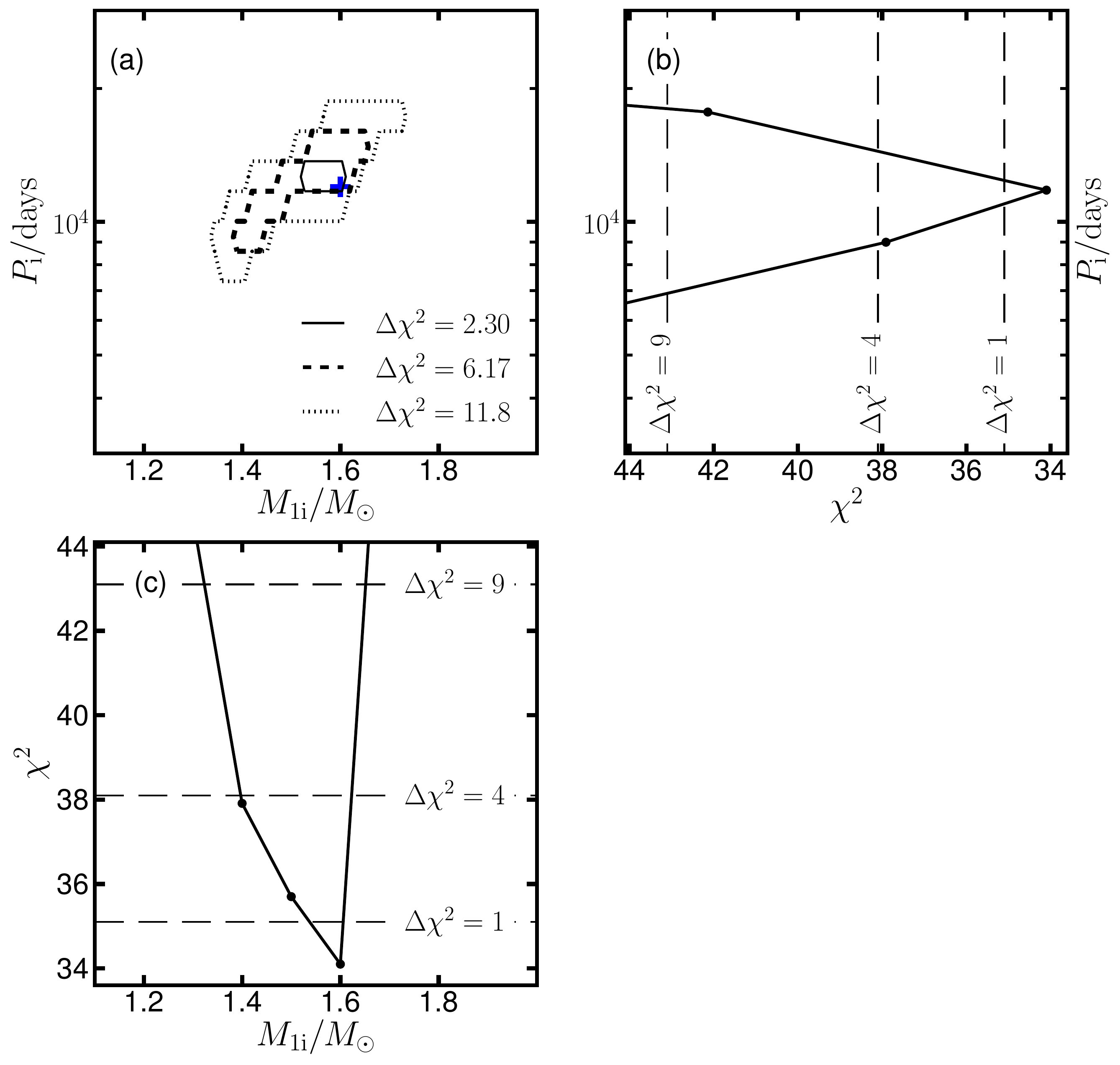}
\caption{Confidence intervals of the input parameters for model star CS29497--034. Panels b and c show the one-dimensional confidence intervals of $\Pin$ and $\Mprim$, respectively. Long-dashed lines indicate the thresholds $\Delta\chisq=1,\,4,\,9$. Panel a shows the two-dimensional confidence intervals. The thresholds $\Delta\chisq=2.30,\,6.17,\,11.8$ are represented as solid, dashed and dotted lines, respectively. The blue plus sign indicates the best model.}
\label{fig:confiCS29497-034}
\end{figure}

\begin{figure}
\centering
\includegraphics[angle=0, width=0.48\textwidth]{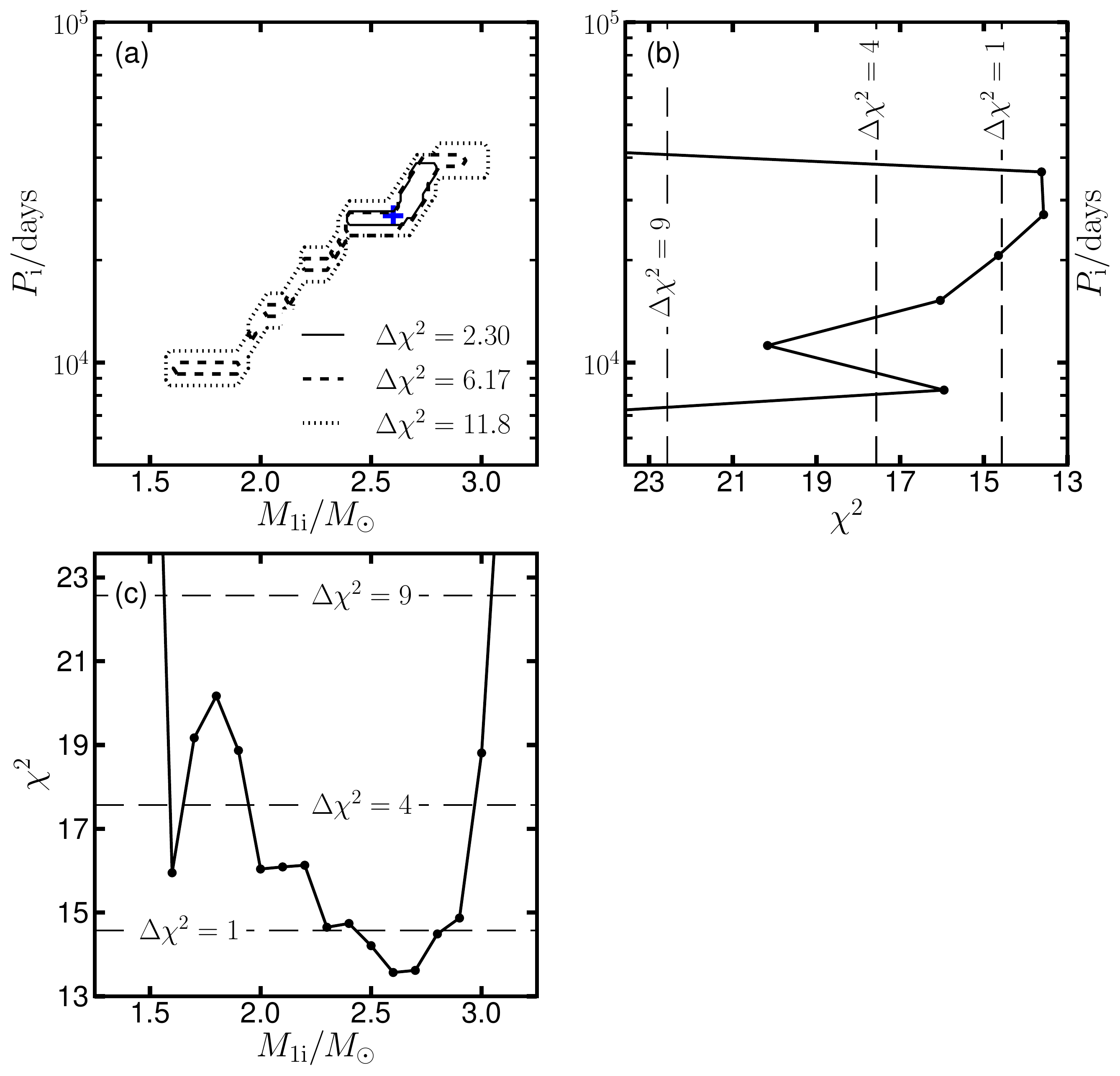}
\caption{
Confidence intervals of the input parameters for model star HD$201626$. The symbols are as in Fig. \ref{fig:confiCS29497-034}.}
\label{fig:confiHD201626}
\end{figure}

Fig. \ref{fig:confiCS29497-034}a shows the two-dimensional confidence regions determined for the initial orbital period, $\Pin$, and primary mass, $\Mprim$, of model star CS29497--034. The one-dimensional confidence intervals of these two parameters are shown in Figs. \ref{fig:confiCS29497-034}b and \ref{fig:confiCS29497-034}c, respectively, in which the same symbols as in Fig. \ref{fig:confiCS22942-019} are used. Fig. \ref{fig:confiHD201626} is the same as Fig. \ref{fig:confiCS29497-034} for star HD201626. To compute the two-dimensional confidence intervals of two input parameters $p$ and $p'$ we follow the same procedure as described in Sect. \ref{method}, but we fix a pair of values $(p,\,p')$ and we let the other two parameters vary. In two dimensions the confidence levels of $68.3\%, 95.4\%$ and $99.7\%$ correspond respectively to the thresholds $\Delta\chisq = 2.30,\, 6.17,\, 11.8$ \cite[][]{NumericalRecipes}. These thresholds are shown in Figs. \ref{fig:confiCS29497-034}a and \ref{fig:confiHD201626}a as solid, dashed and dotted lines, respectively.

The two-dimensional confidence intervals in Fig. \ref{fig:confiCS29497-034}a indicate that there is correlation between the initial orbital period and primary mass of the model systems. The more massive the primary star is, the longer needs to be the initial period to minimise $\chisq$. In Fig. \ref{fig:confiHD201626}a the correlation between the two parameters in the models of star HD201626 is even more clear. This correlation indicates that although the confidence range of $\Mprim$ is rather large, as shown in Table \ref{tab:confi}, for each primary mass in our grid there is essentially only one orbital period at which the model reproduces the observations. 

\vspace{10cm}

%%%%%%%%%%%%%%%%%%%%%%%%%%%%%%%
%

%%%%%%%%%%%%%%%%%%%%%%%%%%%%%%%

\end{appendix}

\end{document}